\ifdefined\twocolumnmode
  \documentclass[journal]{IEEEtran}
\else
  \documentclass[journal,12pt,onecolumn,draftclsnofoot]{IEEEtran}
\fi

\usepackage{amsmath,amssymb,amsfonts}
\usepackage{mathtools}

\usepackage{amsthm}
\usepackage{array}
\usepackage{url}
\usepackage{graphicx}
\usepackage{cite}
\usepackage{algorithm}
\usepackage{algpseudocode}
\usepackage[protrusion=true,expansion=true]{microtype}
\newtheoremstyle{tplain}{4pt}{4pt}{\itshape}{}{\bfseries}{.}{.5em}{}
\newtheoremstyle{tremark}{4pt}{4pt}{\normalfont}{}{\itshape}{.}{.5em}{}
\theoremstyle{tplain}
\newtheorem{theorem}{Theorem}
\newtheorem{lemma}[theorem]{Lemma}
\newtheorem{proposition}[theorem]{Proposition}
\newtheorem{corollary}[theorem]{Corollary}
\newtheorem{definition}[theorem]{Definition}
\theoremstyle{tremark}

\newcommand{\normtwo}[1]{\lVert#1\rVert_2}

\newcommand{\normF}[1]{\lVert#1\rVert_{F}}
\newcommand{\normSp}[2]{\lVert#1\rVert_{S_{#2}}}
\newcommand{\OPT}{\mathrm{OPT}}
\newcommand{\blkdiag}{\mathrm{blkdiag}}

\newcommand{\nnz}{\mathrm{nnz}}
\newcommand{\mc}{m_{\mathrm c}}
\newcommand{\Nrun}{N_{\mathrm{run}}}
\DeclareMathOperator*{\argmin}{arg\,min}

\newcommand{\lestep}[1]{\overset{\mathrm{(#1)}}{\le}}
\newcommand{\eqstep}[1]{\overset{\mathrm{(#1)}}{=}}

\ifdefined\twocolumnmode
  \newcommand{\tcbr}{\\}
  \newenvironment{tcmultline}{\multline}{\endmultline}
  \newenvironment{tcmultline*}{\csname multline*\endcsname}{\csname endmultline*\endcsname}
\else
  \newcommand{\tcbr}{}
  \newenvironment{tcmultline}{\equation}{\endequation}
  \newenvironment{tcmultline*}{\csname equation*\endcsname}{\csname endequation*\endcsname}
\fi

\makeatletter
\newcommand{\itemlbl}[2]{\protected@edef\@currentlabel{\thetheorem(#1)}\label{#2}}
\makeatother
\newcommand{\titem}[2]{\par\smallskip\noindent\textup{(#1)}~\itemlbl{#1}{#2}}

\begin{document}
\bstctlcite{IEEEexample:BSTcontrol}
\title{Communication-Efficient Distributed Sketching for Sparse Linear Regression}

\author{Shashank S,
        Mrityunjoy Chakraborty,~\IEEEmembership{Senior Member,~IEEE,}
        and Soummya Kar,~\IEEEmembership{Fellow,~IEEE}%

\thanks{Shashank S and M. Chakraborty are with the Department of Electronics and Electrical Communication Engineering, Indian Institute of Technology Kharagpur, Kharagpur 721302, India (e-mail: shashanks@kgpian.iitkgp.ac.in, mrityun@ece.iitkgp.ac.in).}%
\thanks{S. Kar is with the Department of Electrical and Computer Engineering, Carnegie Mellon University, Pittsburgh, PA 15213 USA (e-mail: soummyak@andrew.cmu.edu).}%
}

\markboth{}%
{Communication-Efficient Distributed Sketching for Sparse Linear Regression}

\maketitle

\begin{abstract}
This paper considers distributed sketching for $\ell_1$-regularized least squares with data spread across nodes that communicate with a central node, where communication is a bottleneck. We seek minimal communication cost distributed sketching mechanisms, i.e., in which the nodes individually sketch their data prior to communication to the central node, such that, with high probability, the solution of the sketched problem has objective value at most $1+\varepsilon$ times the optimum, for a small $\varepsilon$. The existing baseline, the centralized Gaussian sketch, attains this guarantee and can be implemented across the nodes, but its communication cost scales linearly with the number of nodes. We show that a block-diagonal Gaussian sketch conforming to node data locality, i.e., in which each node sketches its local data and sends the sketch to the central node, attains the same guarantee with a communication cost that, up to lower-order terms, scales with a measure of data heterogeneity across the nodes, which can be sublinear in the number of nodes. From a data storage perspective, a second Gaussian sketch at the central node gives a two-stage sketch with the same guarantee and a sketch size of the same order as that of the centralized sketch, without additional communication. Further, specific large-scale problem scenarios are investigated that demonstrate strict communication and computational advantages of the proposed single-stage distributed and two-stage sketching solutions with respect to the centralized baseline.
\end{abstract}


\section{Introduction}
\label{sec:intro}

\IEEEPARstart{L}{et} $\mathbf A\in\mathbb R^{n\times d}$ be a matrix with columns $\mathbf a_1,\dots,\mathbf a_d$, let $\mathbf b\in\mathbb R^n$ be a vector, and let $\lambda\in(0,1)$. In least absolute shrinkage and selection operator (LASSO) regression~\cite{tibshirani1996}, one solves the $\ell_1$-regularized least-squares problem
\[
\min_{\mathbf x\in\mathbb R^d}F(\mathbf x),\qquad F(\mathbf x):=\normtwo{\mathbf A\mathbf x-\mathbf b}^2+\lambda\|\mathbf x\|_1,
\]
with optimal value $\OPT$. A sketch-and-solve method draws a random matrix $\mathbf S\in\mathbb R^{m\times n}$ with $m\ll n$, where $m$ is the sketch size, and returns
\[
\widetilde{\mathbf x}:=\argmin_{\mathbf x\in\mathbb R^d}\normtwo{\mathbf S(\mathbf A\mathbf x-\mathbf b)}^2+\lambda\|\mathbf x\|_1.
\]
For given $\varepsilon,\delta\in(0,1)$, the goal is $F(\widetilde{\mathbf x})\le(1+O(\varepsilon))\,\OPT$ with probability at least $1-\delta$, with as small a sketch size $m$ as possible. The LASSO objective is not scale-invariant, so some normalization is needed. As in~\cite{mai2023}, we assume $\normtwo{\mathbf A}\le1$ and $\normtwo{\mathbf b}\le1$.

The LASSO is a standard method for sparse regression and variable selection~\cite{tibshirani1996}. In signal processing, it appears as basis pursuit denoising and underlies compressed sensing~\cite{chendonohosaunders1998,donoho2006,candesrombergtao2006}. Sketching reduces the cost of solving such problems at scale~\cite{woodruff2014,pilanciwainwright2015,mai2023}.

Distributed sketching is useful in any setting in which the rows of $[\mathbf A,\mathbf b]$ are acquired or stored at $L$ different nodes, and each node compresses only the rows it holds. Examples include antenna arrays that must compress at each antenna before transmission, and edge nodes that compress before sending data to a cloud~\cite{srinivasa2020}. In sensing, each sensor may observe only part of a signal~\cite{eftekhari2015}. Distributed compressed sensing and imaging systems share this structure~\cite{koep2022}. In all these settings, sending the raw data to a central node is costly, and the central node may not have the capacity to store and process the full data~\cite{charalambidesmazumdar2024}. Each node computes only on its own rows, so the resulting sketch is block-diagonal, and the blocks can be sketched in parallel~\cite{eftekhari2015,srinivasa2020}.

In this paper, the $n$ rows are distributed among $L$ nodes that communicate with a central node, and the goal is to attain the guarantee with minimal communication cost. Node $\ell$ holds the rows $[\mathbf A_\ell,\mathbf b_\ell]\in\mathbb R^{n_\ell\times(d+1)}$. The central node first aggregates the scalars $\normtwo{\mathbf b_\ell}^2$ and broadcasts $\normtwo{\mathbf b}^2$ to all the nodes. Node $\ell$ then estimates $\normtwo{[\mathbf A_\ell,\mathbf b_\ell]}$ from its own data, sets its sketch size $m_\ell$ using this estimate and $\normtwo{\mathbf b_\ell}/\normtwo{\mathbf b}$, samples $\mathbf S_\ell\in\mathbb R^{m_\ell\times n_\ell}$, and sends $\mathbf S_\ell[\mathbf A_\ell,\mathbf b_\ell]$ to the central node. The overall sketch is $\mathbf S=\blkdiag(\mathbf S_1,\dots,\mathbf S_L)$. The central node can then compress the stacked sketches again with a Gaussian sketch $\widetilde{\mathbf S}$, which reduces the sketch size without further communication. We show that this two-stage sketch has a sketch size of the same order as that of the centralized sketch, and that, up to lower-order terms, its communication cost grows with a heterogeneity measure of the data across the nodes and can therefore scale sublinearly with $L$.

\subsection{Existing Work and Research Question}
\label{subsec:priorwork}

Existing work closest to this paper can be divided into three classes: the centralized sketch, block-diagonal sketches, and the distributed implementation of the centralized sketch. We discuss each class below and contrast it with the setup of this paper.

\emph{Centralized sketch.} The work~\cite{mai2023} takes $\mathbf S$ with i.i.d.\ $\mathcal N(0,1/m)$ entries, which we call the centralized sketch. It shows that a sketch size $m=O(\log(d/\delta)/(\varepsilon^2\lambda^2))$ gives the relative-error guarantee $F(\widetilde{\mathbf x})\le(1+O(\varepsilon))\OPT$ for the LASSO with probability at least $1-\delta$~\cite[Thm.~12]{mai2023}, so the dependence on $d$ is only logarithmic. A lower bound in the same work shows that any oblivious sketch needs a sketch size $\Omega(\log(\lambda d)/\lambda^2)$, which matches the upper bound in its dependence on $d$ and $\lambda$~\cite[Thm.~13]{mai2023}. This guarantee assumes that the full data $[\mathbf A,\mathbf b]$ is available at a single node, which is not the case in the setup of this paper.

\emph{Block-diagonal sketches.} The work~\cite{srinivasa2020} takes Gaussian blocks whose sketch sizes are proportional to the coherence parameters of the data blocks~\cite[Eq.~(7)]{srinivasa2020}, and obtains guarantees for approximate matrix multiplication and ridge regression. Its total sketch size matches that of the centralized sketch when an orthonormal basis of the column space of the data is incoherent. Local sketches at the nodes, followed by a second sketch at a coordinator, give $\ell_2$-subspace embeddings for least squares~\cite{charalambidesmazumdar2024}. A subspace embedding of the column space of $[\mathbf A,\mathbf b]$ suffices for the LASSO guarantee but needs a sketch size at least linear in $d$, whereas the centralized sketch of~\cite{mai2023} needs a sketch size only logarithmic in $d$.

\emph{Distributed implementation of the centralized sketch.} The centralized sketch can be applied without gathering the data: each node sketches its own rows with its column block of the centralized sketch, and the central node adds the results (Section~\ref{sec:baseline}). Let $\mc=O(\log(d/\delta)/(\varepsilon^2\lambda^2))$ denote the sketch size of the centralized sketch. Each node then sends $\mc(d+1)$ real numbers, so the total communication cost $L\mc(d+1)$ grows linearly in $L$, regardless of how the data is spread across the nodes.

To our knowledge, none of these results gives a relative-error LASSO guarantee for distributed sketches under a communication bottleneck. This paper addresses the following question.
\begin{quote}
\emph{(Q) Can a distributed sketch attain $F(\widetilde{\mathbf x})\le(1+O(\varepsilon))\,\OPT$ with probability at least $1-\delta$, with a sketch size of the same order as that of the centralized sketch, and with a communication cost that scales sublinearly with $L$, up to lower-order terms?}
\end{quote}

\subsection{Contributions}
\label{subsec:contributions}

For each node $\ell$, let $\eta_\ell:=\tfrac14\normtwo{[\mathbf A_\ell,\mathbf b_\ell]}^2$ and $\beta_\ell:=\normtwo{\mathbf b_\ell}^2/\normtwo{\mathbf b}^2$. We define the \emph{heterogeneity measure} of the data across the nodes as
\[
\nu:=\frac{\sum_{\ell=1}^L\normtwo{[\mathbf A_\ell,\mathbf b_\ell]}^2}{\normtwo{[\mathbf A,\mathbf b]}^2}\in[1,L].
\]
The contributions are as follows.

\begin{enumerate}
\item \emph{Block-diagonal sketch.} We show that a block-diagonal Gaussian sketch $\mathbf S=\blkdiag(\mathbf S_1,\dots,\mathbf S_L)$, in which node $\ell$ chooses its sketch size $m_\ell$ from $\eta_\ell$ and $\beta_\ell$, attains $F(\widetilde{\mathbf x})\le(1+O(\varepsilon))\OPT$ with probability at least $1-\delta$ (Theorem~\ref{thm:main}). The total sketch size is
\[
m=O\bigg(\frac{\nu}{\varepsilon^2\lambda^2}\log\frac{d\log(2/(\varepsilon\lambda))}{\delta}\bigg),
\]
which scales with $\nu$ rather than with $L$. When $\nu=O(1)$, it matches the sketch size $O(\log(d/\delta)/(\varepsilon^2\lambda^2))$ of the centralized sketch~\cite[Thm.~12]{mai2023} in order.

\item \emph{Two-stage sketch.} We show that a second sketch at the central node removes the dependence of the final sketch size on $\nu$. The nodes apply a block-diagonal sketch $\widehat{\mathbf S}$ as in item~1, with slightly tighter parameters. The central node then applies a Gaussian sketch $\widetilde{\mathbf S}\in\mathbb R^{\widetilde m\times m}$, drawn independently of $\widehat{\mathbf S}$, to the stacked local sketches. The two-stage sketch $\widetilde{\mathbf S}\widehat{\mathbf S}$ attains $F(\widetilde{\mathbf x})\le(1+O(\varepsilon))\OPT$ with probability at least $1-\delta$ (Theorem~\ref{thm:twostage}), with sketch size
\[
\widetilde m=O\bigg(\frac{1}{\varepsilon^2\lambda^2}\log\frac{d\log(2/(\varepsilon\lambda))}{\delta}\bigg),
\]
independent of $L$ and $\nu$. This matches the sketch size of the centralized sketch~\cite[Thm.~12]{mai2023} in order. Since the second stage is applied at the central node, it needs no communication. The nodes send only their local sketches, of $m(d+1)$ entries in total, together with the scalars $\normtwo{\mathbf b_\ell}^2$. Up to lower-order terms, the communication cost therefore scales with $\nu$ rather than with $L$, and it is a fraction of order $\nu/L$ of the cost $L\mc(d+1)$ of the distributed implementation of the centralized sketch.

\item \emph{Computation cost and an example.} We also compare the computation costs, and show that, under mild conditions, the two-stage sketch requires less computation than the distributed implementation of the centralized sketch (Section~\ref{sec:cost}). We then give an example in which $\nu$ is bounded by a constant: for Gaussian data split equally across the nodes, $\nu\le25/9$ for all sufficiently large $L$, almost surely (Proposition~\ref{prop:gaussian} and Corollary~\ref{cor:asymptotic}). In this example, the communication cost is reduced by a factor of order $L$.
\end{enumerate}
\subsection{Related Work}
\label{subsec:related}

This subsection places the paper within two lines of work.

\emph{Sketching for regularized and constrained regression.} For ridge regression, the statistical dimension replaces the rank in the required sketch size~\cite{avronclarksonwoodruff2017}. Iterative sketching solves ridge regression to high accuracy~\cite{chowdhury2018,kachamwoodruff2022}. For least squares over a convex set, the geometry of the set governs the sketch size~\cite{pilanciwainwright2015}. Convex and nonconvex regularizers admit guarantees on the error of the sketched solution~\cite{yangli2025}. Analyses tailored to a specific objective can require a much smaller sketch size than a subspace embedding~\cite{pilanciwainwright2015,mai2023}. Sketches are also known for logistic regression and for the $\ell_1$ loss~\cite{munteanu2021,munteanu2023}, and for sparse regression under several losses~\cite{mai2023}.

\emph{Block-diagonal and distributed sketching.} Block-diagonal random matrices satisfy the restricted isometry property for sparse signals~\cite{eftekhari2015}, and this property extends to group-sparse signals~\cite{koep2022}. Structured measurement matrices generally need more measurements than unstructured ones~\cite{eftekhari2015}. These analyses rest on bounds for suprema of chaos processes~\cite{krahmermendelsonrauhut2014,dirksen2015}. In distributed regression, sketching enables privacy, straggler resilience and asynchronous updates~\cite{bartanpilanci2020}. It also reduces communication cost and memory footprint in distributed least squares~\cite{garg2024}. In federated learning, it compresses model updates~\cite{haddadpour2020,rothchild2020}. These methods average full-size sketches or compress model updates by sketching. They do not use block-diagonal sketches of the data.

\subsection{Organization}
\label{subsec:organization}

Section~\ref{sec:setup} states the problem and the key assumptions, recalls the guarantee for the centralized sketch~\cite[Thm.~12]{mai2023} and the localized sketching guarantee~\cite[Lem.~1]{srinivasa2020}, and describes the baseline, the distributed implementation of the centralized sketch (Section~\ref{sec:baseline}). It then states Conditions C1 and C2, the main results, Theorems~\ref{thm:partI}--\ref{thm:twostage}, and the communication cost of the two-stage sketch (Section~\ref{sec:mainresult}). It ends with a discussion of the heterogeneity measure $\nu$, including an example with Gaussian data in which $\nu$ is bounded by a constant (Section~\ref{sec:gaussian}), and a comparison of the computation cost of the two-stage sketch with that of the distributed implementation of the centralized sketch (Section~\ref{sec:cost}). Section~\ref{sec:partI} proves the main results and shows how each node estimates $\eta_\ell$ from its own data (Section~\ref{sec:etaest}). Section~\ref{sec:conclusion} concludes. Appendix~\ref{app:shell} proves Lemma~\ref{lem:shell}, which completes the proof of Theorem~\ref{thm:ma}, and Appendix~\ref{app:gamma2} proves Lemma~\ref{lem:gamma2contract}.

\subsection{Notation}
\label{subsec:notation-conventions}

Bold lowercase letters denote vectors ($\mathbf a_i,\mathbf b,\mathbf x$), and bold uppercase letters denote matrices ($\mathbf A,\mathbf S$). For a vector $\mathbf v$, $\normtwo{\mathbf v}$, $\|\mathbf v\|_1$ and $\|\mathbf v\|_\infty$ are its Euclidean, $\ell_1$ and $\ell_\infty$ norms, and $\langle\cdot,\cdot\rangle$ is the Euclidean inner product. For a matrix $\mathbf M$, $\normtwo{\mathbf M}$ is its spectral norm, $\normF{\mathbf M}$ its Frobenius norm, $\normSp{\mathbf M}{p}$ its Schatten-$p$ norm, $\|\mathbf M\|_{\max}:=\max_{i,j}|M_{ij}|$, where $M_{ij}$ are its entries, $\sigma_{\max}(\mathbf M)$ and $\sigma_{\min}(\mathbf M)$ its largest and smallest singular values, $\operatorname{tr}(\mathbf M)$ its trace, and $\nnz(\mathbf M)$ its number of nonzero entries. The Kronecker product is $\otimes$, $\mathrm{conv}$ denotes the convex hull, and $\mathbf 1_E$ is the indicator of an event $E$. For $\mathbf A\in\mathbb R^{n\times d}$ and $\mathbf b\in\mathbb R^n$, $[\mathbf A,\mathbf b]\in\mathbb R^{n\times(d+1)}$ is the matrix obtained by appending $\mathbf b$ to $\mathbf A$ as a last column, and $\blkdiag(\mathbf S_1,\dots,\mathbf S_L)$ is the block-diagonal matrix with diagonal blocks $\mathbf S_1,\dots,\mathbf S_L$. For a scalar random variable $\xi$, $\|\xi\|_{\psi_2}:=\inf\{t>0\mid\mathbb E\exp(\xi^2/t^2)\le2\}$ is its sub-Gaussian norm. For a random vector $\boldsymbol\xi=(\xi_1,\dots,\xi_n)$, $\|\boldsymbol\xi\|_{\psi_2}:=\max_i\|\xi_i\|_{\psi_2}$. For $x\in\mathbb R$, $\lceil x\rceil$ is the smallest integer greater than or equal to $x$. The logarithm $\log$ is natural unless a base is shown. For nonnegative $f$ and $g$, $f=O(g)$, or equivalently $g=\Omega(f)$, means $f\le c\,g$ for an absolute constant $c$, and $f=\Theta(g)$ means $f=O(g)$ and $g=O(f)$.
\section{Problem Setup and Main Results}
\label{sec:setup}

This section states the problem, the two existing results we use, the distributed implementation of the centralized sketch, two conditions on the sketch, and the main results. It ends with a discussion of the heterogeneity measure $\nu$ and of the computation cost.

\subsection{Problem Setup and Key Assumptions}
\label{sec:notation}

\textbf{Problem Setup.} Let $\mathbf A\in\mathbb R^{n\times d}$ have columns $\mathbf a_1,\dots,\mathbf a_d\in\mathbb R^n$, let $\mathbf b\in\mathbb R^n$, and let $\mathbf S\in\mathbb R^{m\times n}$ be a sketching matrix. Define the objective function, its minimizer, and the minimizer of the sketched problem:
\begin{align*}
F(\mathbf x)&:=\normtwo{\mathbf A\mathbf x-\mathbf b}^2+\lambda\|\mathbf x\|_1,\\
\mathbf x^*&:=\argmin_{\mathbf x}F(\mathbf x),\\
\widetilde{\mathbf x}&:=\argmin_{\mathbf x}\normtwo{\mathbf S(\mathbf A\mathbf x-\mathbf b)}^2+\lambda\|\mathbf x\|_1.
\end{align*}
If a minimizer is not unique, $\mathbf x^*$ and $\widetilde{\mathbf x}$ denote any fixed minimizers. Write $\OPT:=F(\mathbf x^*)$ for the optimal objective value.

Recall that $[\mathbf A,\mathbf b]\in\mathbb R^{n\times(d+1)}$ has columns $\mathbf a_1,\dots,\mathbf a_d,\mathbf b$. For block-diagonal sketches, the $n$ rows are partitioned into $L$ consecutive blocks of sizes $n_1,\dots,n_L$, with $\sum_\ell n_\ell=n$. Block $\ell$ consists of the rows $[\mathbf A_\ell,\mathbf b_\ell]\in\mathbb R^{n_\ell\times(d+1)}$ of $[\mathbf A,\mathbf b]$. We assume $[\mathbf A_\ell,\mathbf b_\ell]\ne\mathbf 0$ for every $\ell$, since a zero block can be removed without changing the problem. Recall from Section~\ref{subsec:contributions} that $\eta_\ell=\tfrac14\normtwo{[\mathbf A_\ell,\mathbf b_\ell]}^2$, $\beta_\ell=\normtwo{\mathbf b_\ell}^2/\normtwo{\mathbf b}^2$ and
\[
\nu=\frac{\sum_{\ell=1}^L\normtwo{[\mathbf A_\ell,\mathbf b_\ell]}^2}{\normtwo{[\mathbf A,\mathbf b]}^2}.
\]
We set $\beta_\ell:=0$ when $\mathbf b=\mathbf 0$, and let $\eta_\Sigma:=\sum_{\ell=1}^L\eta_\ell$, so that $\eta_\Sigma=\nu\normtwo{[\mathbf A,\mathbf b]}^2/4$. Section~\ref{sec:rowcount} shows that $1\le\nu\le L$.

\textbf{Key Assumptions.}
\[
\normtwo{\mathbf A}\le1,\qquad \normtwo{\mathbf b}\le1,\qquad \lambda\in(0,1),\qquad \varepsilon,\delta\in(0,1).
\]
Since $\mathbf a_i=\mathbf A\mathbf e_i$ for each standard basis vector $\mathbf e_i\in\mathbb R^d$, we have $\normtwo{\mathbf a_i}\le\normtwo{\mathbf A}\le1$ for $1\le i\le d$. These assumptions give $\OPT\le F(\mathbf 0)=\normtwo{\mathbf b}^2\le1$.

\subsection{Existing Results}
\label{sec:preliminaries}

In this subsection, we state two existing results: the guarantee for the centralized sketch~\cite[Thm.~12]{mai2023} and the localized sketching guarantee~\cite[Lem.~1]{srinivasa2020}.

The first result is the guarantee for the centralized sketch.

\begin{theorem}[{Centralized LASSO sketch~\cite[Thm.~12]{mai2023}}]
\label{thm:mai12}
Let $\mathbf A\in\mathbb R^{n\times d}$ and $\mathbf b\in\mathbb R^n$ satisfy $\normtwo{\mathbf A}\le1$ and $\normtwo{\mathbf b}\le1$, and let $\lambda,\varepsilon,\delta\in(0,1)$. Let $\mathbf S\in\mathbb R^{m\times n}$ have i.i.d.\ $\mathcal N(0,1/m)$ entries, with $m=O\big(\log(d/\delta)/(\varepsilon^2\lambda^2)\big)$. Then, with probability at least $1-\delta$, the minimizer $\widetilde{\mathbf x}:=\argmin_{\mathbf x\in\mathbb R^d}\normtwo{\mathbf S(\mathbf A\mathbf x-\mathbf b)}^2+\lambda\|\mathbf x\|_1$ satisfies
\[
F(\widetilde{\mathbf x})\le(1+\varepsilon)\,\OPT.
\]
\end{theorem}

The second result concerns block-diagonal Gaussian sketches. We use the following definitions from~\cite{srinivasa2020}. For $\mathbf U\in\mathbb R^{n\times r}$ with orthonormal columns, let $\mathbf U_\ell\in\mathbb R^{n_\ell\times r}$ be its restriction to the rows of block $\ell$. The coherence of block $\ell$ is~\cite[Sec.~2.1]{srinivasa2020}
\[
\Gamma(\mathbf U_\ell):=\min\big\{n_\ell\,\|\mathbf U_\ell\|_{\max}^2,\ \normtwo{\mathbf U_\ell}^2\big\}.
\]
The work~\cite{srinivasa2020} defines $\Gamma$ for blocks of a common length~\cite[Sec.~2.1]{srinivasa2020} and notes that its results extend to blocks of different sizes~\cite[Sec.~1]{srinivasa2020}.
\begin{lemma}[{Localized sketching guarantee~\cite[Lem.~1]{srinivasa2020}}]
\label{lem:srinivasa}
Let $\mathbf U\in\mathbb R^{n\times r}$ have orthonormal columns. Let $\mathbf S=\blkdiag(\mathbf S_1,\dots,\mathbf S_L)$, where the entries of $\mathbf S_\ell\in\mathbb R^{M_\ell\times n_\ell}$ are i.i.d.\ $\mathcal N(0,1/M_\ell)$, independent across blocks, and $M_\ell=M_0\,\Gamma(\mathbf U_\ell)$. Let $\mathbf K:=(\mathbf S\mathbf U)^\top(\mathbf S\mathbf U)-\mathbf I_r$. There is an absolute constant $c_1$ such that, for every $\delta\in(0,1)$,
\[
\mathbb P\Big(\normtwo{\mathbf K}\le c_1\sqrt{r\log(2/\delta)/M_0}\Big)\ge1-\delta.
\]
In particular, for every $\varepsilon'\in(0,1)$, if $M_0\ge c_1^2\,r\log(2/\delta)/\varepsilon'^2$, then $\mathbb P(\normtwo{\mathbf K}\le\varepsilon')\ge1-\delta$.
\end{lemma}

\subsection{Baseline: Distributed Implementation of the Centralized Sketch}
\label{sec:baseline}

This subsection describes the baseline against which the distributed sketches of this paper are compared. Let $\mathbf G=[\mathbf G_1,\dots,\mathbf G_L]\in\mathbb R^{\mc\times n}$ have i.i.d.\ $\mathcal N(0,1/\mc)$ entries, with column blocks $\mathbf G_\ell\in\mathbb R^{\mc\times n_\ell}$ conformal with the row blocks of $[\mathbf A,\mathbf b]$. Node $\ell$ draws $\mathbf G_\ell$ and sends $\mathbf G_\ell[\mathbf A_\ell,\mathbf b_\ell]$ to the central node. The central node computes $\sum_\ell\mathbf G_\ell[\mathbf A_\ell,\mathbf b_\ell]=\mathbf G[\mathbf A,\mathbf b]$, which is the data sketched by the centralized sketch $\mathbf G$. Theorem~\ref{thm:mai12} therefore applies with $\mc=O(\log(d/\delta)/(\varepsilon^2\lambda^2))$. Each node sends $\mc(d+1)$ real numbers, so the total communication cost is $L\,\mc(d+1)$ real numbers. It grows linearly in $L$ and does not depend on how the data is spread across the nodes. In contrast, up to lower-order terms, the total communication cost of the block-diagonal sketch of Section~\ref{sec:mainresult} scales with $\nu$.

\subsection{Main Results}
\label{sec:mainresult}

This subsection states the main results. Their proofs are in Section~\ref{sec:partI}.

\subsubsection{Sufficient Conditions}
We first state two conditions on $\mathbf S$ under which the sketched LASSO attains the relative-error guarantee. Let
\[
T:=\mathrm{conv}\{\pm\mathbf a_1,\dots,\pm\mathbf a_d,\pm\mathbf b\}\subset\mathbb R^n,
\]
the convex hull of $2d+2$ points. Each point has norm at most $1$ by Section~\ref{sec:notation}.

\noindent\textbf{C1$(\varepsilon,\lambda)$.} For every $\mathbf z\in T$,
\[
\big|\normtwo{\mathbf S\mathbf z}-\normtwo{\mathbf z}\big|\le\varepsilon\lambda.
\]

\noindent\textbf{C2$(\lambda)$.}
\[
\big|\normtwo{\mathbf S\mathbf b}^2-\normtwo{\mathbf b}^2\big|\le\tfrac{\lambda}{3}\normtwo{\mathbf b}^2.
\]

The first result shows that C1$(\varepsilon,\lambda)$ and C2$(\lambda)$ suffice for the LASSO guarantee.

\begin{theorem}
\label{thm:partI}
Let $\mathbf A$, $\mathbf b$ and $\lambda$ be as in Section~\ref{sec:notation}, suppose that $\mathbf S$ satisfies C1$(\varepsilon,\lambda)$ and C2$(\lambda)$, and let $\widetilde{\mathbf x}:=\argmin_{\mathbf x\in\mathbb R^d}\normtwo{\mathbf S(\mathbf A\mathbf x-\mathbf b)}^2+\lambda\|\mathbf x\|_1$. Then
\begin{tcmultline*}
F(\widetilde{\mathbf x})\le\big(1+(121+918\varepsilon+867\varepsilon^2)\,\varepsilon\big)\,\OPT \tcbr
\le(1+1906\,\varepsilon)\,\OPT=\big(1+O(\varepsilon)\big)\,\OPT.
\end{tcmultline*}
\end{theorem}

Its proof is in Section~\ref{sec:partI-distortion}.

\subsubsection{Block-Diagonal Sketch}
The next result gives the sketch sizes under which a block-diagonal Gaussian sketch satisfies each of C1$(\varepsilon,\lambda)$ and C2$(\lambda)$.

\begin{theorem}
\label{thm:mamb}
There exist absolute constants $c_2$, $c_3$ and $c_4$ such that the following holds. Let $\mathbf A$, $\mathbf b$, $\lambda$ and the block partition be as in Section~\ref{sec:notation}, and let $\varepsilon,\delta\in(0,1)$. Let $\mathbf S=\blkdiag(\mathbf S_1,\dots,\mathbf S_L)$, where the entries of $\mathbf S_\ell\in\mathbb R^{m_\ell\times n_\ell}$ are i.i.d.\ $\mathcal N(0,1/m_\ell)$, independent across blocks. Define
\begin{tcmultline}
m_1(\varepsilon,\lambda,\delta):=\frac{c_3}{\varepsilon^2\lambda^2}\max\Big\{2c_2^2\log\big(2(d+1)\big), \tcbr
\log\frac1\delta,\ \log\log\frac{2}{\varepsilon\lambda}\Big\}, \label{eq:mafun}
\end{tcmultline}
\begin{equation}
m_2(\lambda,\delta):=\frac{9c_4\log(2/\delta)}{\lambda^2}. \label{eq:mbfun}
\end{equation}
\titem{i}{thm:ma}If $m_\ell\ge m_1(\varepsilon,\lambda,\delta)\,\eta_\ell$ for every $\ell$, then $\mathbf S$ satisfies C1$(\varepsilon,\lambda)$ with probability at least $1-\delta$.
\titem{ii}{thm:mb}If $m_\ell\ge m_2(\lambda,\delta)\,\beta_\ell$ for every $\ell$, then $\mathbf S$ satisfies C2$(\lambda)$ with probability at least $1-\delta$.
\end{theorem}

Its proof is in Section~\ref{sec:partI-chaining}. The orders of the two functions are
\begin{align*}
m_1(\varepsilon,\lambda,\delta)&=O\bigg(\frac{1}{\varepsilon^2\lambda^2}\log\frac{d\log(2/(\varepsilon\lambda))}{\delta}\bigg),\\
m_2(\lambda,\delta)&=O\Big(\frac{\log(2/\delta)}{\lambda^2}\Big).
\end{align*}
The function $m_1$ also depends on $d$, which we suppress in the notation; in Section~\ref{sec:gaussian}, it is evaluated at $d=d_L$.

The next result combines Theorems~\ref{thm:partI} and~\ref{thm:mamb}. Each of the two conditions is satisfied with probability at least $1-\delta/2$.

\begin{theorem}
\label{thm:main}
Let $\mathbf A$, $\mathbf b$, $\lambda$ and the block partition be as in Section~\ref{sec:notation}, and let $\varepsilon,\delta\in(0,1)$. Let $\mathbf S=\blkdiag(\mathbf S_1,\dots,\mathbf S_L)$, where the entries of $\mathbf S_\ell\in\mathbb R^{m_\ell\times n_\ell}$ are i.i.d.\ $\mathcal N(0,1/m_\ell)$, independent across blocks, and
\begin{tcmultline}
m_\ell\ge\max\big\{m_1(\varepsilon,\lambda,\delta/2)\,\eta_\ell,\ m_2(\lambda,\delta/2)\,\beta_\ell\big\}, \tcbr
\qquad\ell=1,\dots,L. \label{eq:allocation-main}
\end{tcmultline}
Then, with probability at least $1-\delta$, $\mathbf S$ satisfies C1$(\varepsilon,\lambda)$ and C2$(\lambda)$. Consequently, by Theorem~\ref{thm:partI}, the solution $\widetilde{\mathbf x}$ satisfies $F(\widetilde{\mathbf x})\le(1+O(\varepsilon))\,\OPT$. In particular, $m_\ell=\big\lceil\max\{m_1(\varepsilon,\lambda,\delta/2)\,\eta_\ell,\ m_2(\lambda,\delta/2)\,\beta_\ell\}\big\rceil$ suffices, and gives the total sketch size $m=\sum_{\ell=1}^Lm_\ell=O\big(m_1(\varepsilon,\lambda,\delta/2)\,\eta_\Sigma+m_2(\lambda,\delta/2)\big)$.
\end{theorem}

Its proof is in Section~\ref{sec:mainproof}. When $\nu=O(1)$, the sketch size has the same dependence on $d$ and $\delta$ as that of the centralized sketch in Theorem~\ref{thm:mai12}.

The information that node $\ell$ needs to apply its local sketch of Theorem~\ref{thm:main} in a distributed setting is $\normtwo{[\mathbf A_\ell,\mathbf b_\ell]}$, $\normtwo{\mathbf b_\ell}$ and $\normtwo{\mathbf b}$. The first two are computed locally, without any communication. For the last, the central node aggregates the scalars $\normtwo{\mathbf b_\ell}^2$ from all nodes and broadcasts $\normtwo{\mathbf b}^2$.

\subsubsection{Two-Stage Sketch}
In the two-stage sketch, the nodes apply a block-diagonal sketch $\widehat{\mathbf S}=\blkdiag(\widehat{\mathbf S}_1,\dots,\widehat{\mathbf S}_L)$ as in Theorem~\ref{thm:main}. By Theorem~\ref{thm:main} and Section~\ref{sec:rowcount}, the stack of locally sketched data $\widehat{\mathbf S}[\mathbf A,\mathbf b]\in\mathbb R^{m\times(d+1)}$ at the central node can have a sketch size of order up to $\nu$ times that of the centralized sketch in Theorem~\ref{thm:mai12}. To reduce it to that of the centralized sketch, the central node applies a Gaussian sketch $\widetilde{\mathbf S}\in\mathbb R^{\widetilde m\times m}$, drawn independently of $\widehat{\mathbf S}$, to obtain $\widetilde{\mathbf S}\widehat{\mathbf S}[\mathbf A,\mathbf b]\in\mathbb R^{\widetilde m\times(d+1)}$. The two-stage sketch is $\mathbf S:=\widetilde{\mathbf S}\widehat{\mathbf S}$. The next result gives the sketch size of $\widetilde{\mathbf S}$ under which $\widetilde{\mathbf S}\widehat{\mathbf S}$ satisfies C1$(\varepsilon,\lambda)$ and C2$(\lambda)$, provided that $\widehat{\mathbf S}$ satisfies C1$(\varepsilon/2,\lambda)$ and C2$(3\lambda/7)$.

\begin{theorem}
\label{thm:transfer}
There exist absolute constants $c_5$ and $c_6$ such that the following holds. Let $\mathbf A$, $\mathbf b$ and $\lambda$ be as in Section~\ref{sec:notation}, and let $\varepsilon,\delta\in(0,1)$. Let $\widehat{\mathbf S}\in\mathbb R^{m\times n}$, and let $\widetilde{\mathbf S}\in\mathbb R^{\widetilde m\times m}$ have i.i.d.\ $\mathcal N(0,1/\widetilde m)$ entries, independent of $\widehat{\mathbf S}$. Define
\begin{equation}
\widetilde m_1(\varepsilon,\lambda,\delta):=\frac{c_5}{\varepsilon^2\lambda^2}\log\frac{(2d+2)\log(2/(\varepsilon\lambda))}{\delta}, \label{eq:mtildea}
\end{equation}
\begin{equation}
\widetilde m_2(\lambda,\delta):=\frac{c_6}{\lambda^2}\log\frac1\delta. \label{eq:mtildeb}
\end{equation}
\titem{i}{thm:transfer-a}If $\widehat{\mathbf S}$ satisfies C1$(\varepsilon/2,\lambda)$ and $\widetilde m\ge\widetilde m_1(\varepsilon,\lambda,\delta)$, then, conditionally on $\widehat{\mathbf S}$, the matrix $\widetilde{\mathbf S}\widehat{\mathbf S}$ satisfies C1$(\varepsilon,\lambda)$ with probability at least $1-\delta$.
\titem{ii}{thm:transfer-b}If $\widehat{\mathbf S}$ satisfies C2$(3\lambda/7)$ and $\widetilde m\ge\widetilde m_2(\lambda,\delta)$, then, conditionally on $\widehat{\mathbf S}$, the matrix $\widetilde{\mathbf S}\widehat{\mathbf S}$ satisfies C2$(\lambda)$ with probability at least $1-\delta$.
\end{theorem}

Its proof is in Section~\ref{sec:twostage}. The next theorem combines Theorems~\ref{thm:partI}, \ref{thm:mamb} and~\ref{thm:transfer}.

\begin{theorem}
\label{thm:twostage}
Let $\mathbf A$, $\mathbf b$, $\lambda$ and the block partition be as in Section~\ref{sec:notation}, and let $\varepsilon,\delta\in(0,1)$. Let $\widehat{\mathbf S}=\blkdiag(\widehat{\mathbf S}_1,\dots,\widehat{\mathbf S}_L)$ be as in Theorem~\ref{thm:main}, but with
\begin{tcmultline*}
m_\ell=\Big\lceil\max\Big\{m_1(\varepsilon/2,\lambda,\delta/4)\,\eta_\ell, \tcbr
m_2(3\lambda/7,\delta/4)\,\beta_\ell\Big\}\Big\rceil
\end{tcmultline*}
for $\ell=1,\dots,L$, let $\widetilde{\mathbf S}$ be as in Theorem~\ref{thm:transfer} with $\widetilde m=\big\lceil\max\{\widetilde m_1(\varepsilon,\lambda,\delta/4),\ \widetilde m_2(\lambda,\delta/4)\}\big\rceil$, and let $\mathbf S:=\widetilde{\mathbf S}\widehat{\mathbf S}$. Then, with probability at least $1-\delta$, $\mathbf S$ satisfies C1$(\varepsilon,\lambda)$ and C2$(\lambda)$, and the minimizer $\widetilde{\mathbf x}:=\argmin_{\mathbf x\in\mathbb R^d}\normtwo{\mathbf S(\mathbf A\mathbf x-\mathbf b)}^2+\lambda\|\mathbf x\|_1$ satisfies
\[
F(\widetilde{\mathbf x})\le(1+1906\,\varepsilon)\,\OPT=(1+O(\varepsilon))\,\OPT.
\]
The sketch size satisfies
\[
\widetilde m=O\bigg(\frac{1}{\varepsilon^2\lambda^2}\log\frac{d\log(2/(\varepsilon\lambda))}{\delta}\bigg),
\]
independent of $L$.
\end{theorem}

Its proof is in Section~\ref{sec:twostageproof}. The sketch size $\widetilde m$ has the same dependence on $d$ and $\delta$ as the sketch size $O(\log(d/\delta)/(\varepsilon^2\lambda^2))$ of the centralized sketch in Theorem~\ref{thm:mai12}. Algorithm~\ref{alg:twostage} summarizes the resulting protocol.

\begin{algorithm}[!t]
\caption{Two-stage distributed sketch. In Step~2, $\eta_\ell$ may be replaced by the estimate $\widehat\eta_\ell$ of Lemma~\ref{lem:etaest} (see Section~\ref{sec:etaest}).}
\label{alg:twostage}
\begin{algorithmic}[1]
\Statex \textbf{Input:} node $\ell$ holds $[\mathbf A_\ell,\mathbf b_\ell]$, $\ell=1,\dots,L$; parameters $\varepsilon,\lambda,\delta\in(0,1)$.
\State Each node $\ell$ sends $\normtwo{\mathbf b_\ell}^2$ to the central node, which broadcasts $\normtwo{\mathbf b}^2=\sum_{\ell=1}^L\normtwo{\mathbf b_\ell}^2$.
\State Each node $\ell$ computes $\eta_\ell=\tfrac14\normtwo{[\mathbf A_\ell,\mathbf b_\ell]}^2$ and $\beta_\ell=\normtwo{\mathbf b_\ell}^2/\normtwo{\mathbf b}^2$, and sets $m_\ell:=\big\lceil\max\{m_1(\varepsilon/2,\lambda,\delta/4)\,\eta_\ell,\ m_2(3\lambda/7,\delta/4)\,\beta_\ell\}\big\rceil$.
\State Each node $\ell$ draws $\widehat{\mathbf S}_\ell\in\mathbb R^{m_\ell\times n_\ell}$ with i.i.d.\ $\mathcal N(0,1/m_\ell)$ entries and sends $\widehat{\mathbf S}_\ell[\mathbf A_\ell,\mathbf b_\ell]$ to the central node.
\State The central node stacks the received matrices into $\widehat{\mathbf S}[\mathbf A,\mathbf b]\in\mathbb R^{m\times(d+1)}$, draws $\widetilde{\mathbf S}\in\mathbb R^{\widetilde m\times m}$ as in Theorem~\ref{thm:twostage}, and forms $\mathbf S[\mathbf A,\mathbf b]=\widetilde{\mathbf S}\widehat{\mathbf S}[\mathbf A,\mathbf b]$.
\State The central node returns $\widetilde{\mathbf x}:=\argmin_{\mathbf x\in\mathbb R^d}\normtwo{\mathbf S(\mathbf A\mathbf x-\mathbf b)}^2+\lambda\|\mathbf x\|_1$.
\end{algorithmic}
\end{algorithm}

\subsubsection{Communication Cost}
In Algorithm~\ref{alg:twostage}, node $\ell$ sends the scalar $\normtwo{\mathbf b_\ell}^2$ and the matrix $\widehat{\mathbf S}_\ell[\mathbf A_\ell,\mathbf b_\ell]\in\mathbb R^{m_\ell\times(d+1)}$ to the central node, and the central node broadcasts $\normtwo{\mathbf b}^2$. The second stage is applied at the central node and needs no communication. The total communication is therefore $m(d+1)$ real numbers, together with $L$ scalars and one broadcast. With $m_\ell$ as in Algorithm~\ref{alg:twostage}, $m=O\big(\nu m_1(\varepsilon/2,\lambda,\delta/4)\,+m_2(3\lambda/7,\delta/4)\big)$, so the communication cost scales with $\nu$, up to lower-order terms. The $L$ scalars and the broadcast add $L+1$ real numbers. This cost is of lower order, does not depend on $\varepsilon$, $\lambda$, $d$ or $\delta$, and is dominated by the communication cost of the baseline.

\subsection{Discussion of the Main Results}
\label{sec:rowcount}

The total sketch size in Theorem~\ref{thm:main} has two terms. The second term, $m_2(\lambda,\delta/2)=O(\log(2/\delta)/\lambda^2)$, has no factor $1/\varepsilon^2$ and does not depend on $d$, on $\normtwo{\mathbf b}$ or on the partition. Since the row blocks are disjoint, for every $\mathbf y\in\mathbb R^{d+1}$,
\begin{equation}
\normtwo{[\mathbf A,\mathbf b]\mathbf y}^2=\sum_{\ell=1}^L\normtwo{[\mathbf A_\ell,\mathbf b_\ell]\mathbf y}^2. \label{eq:disjoint}
\end{equation}
By~\eqref{eq:disjoint}, $\normtwo{[\mathbf A_\ell,\mathbf b_\ell]}\le\normtwo{[\mathbf A,\mathbf b]}$ for every $\ell$, which gives $\nu\le L$. Applying~\eqref{eq:disjoint} at a leading right singular vector of $[\mathbf A,\mathbf b]$ gives $\nu\ge1$. Since $\eta_\Sigma=\nu\normtwo{[\mathbf A,\mathbf b]}^2/4$ and $\normtwo{[\mathbf A,\mathbf b]}^2\le2$(shown in Section~\ref{sec:cost}), we have $\eta_\Sigma\le\nu/2$, and the total sketch size in Theorem~\ref{thm:main} satisfies
\[
m=O\big(\nu\,m_1(\varepsilon,\lambda,\delta/2)+m_2(\lambda,\delta/2)\big),\qquad1\le\nu\le L.
\]
Thus $\nu$ determines how the sketch size of the block-diagonal sketch, and hence the communication cost of the two-stage sketch, grows as the data is split over more nodes. In the worst case, $\nu=L$, and the bound of Theorem~\ref{thm:main} is then of order $L$ times the sketch size of the centralized sketch. One such case is when the row spaces of the blocks $[\mathbf A_\ell,\mathbf b_\ell]$ are mutually orthogonal and the blocks have equal spectral norms. Then $\normtwo{[\mathbf A,\mathbf b]}^2=\max_\ell\normtwo{[\mathbf A_\ell,\mathbf b_\ell]}^2$, so $\nu=L$. At the other extreme, if all the blocks are equal, that is, $[\mathbf A_\ell,\mathbf b_\ell]=[\mathbf A_1,\mathbf b_1]$ for every $\ell$, then $\normtwo{[\mathbf A,\mathbf b]}^2=L\normtwo{[\mathbf A_1,\mathbf b_1]}^2$, so $\nu=1$. More generally, $\nu=1$ whenever the blocks share a leading right singular vector.

Section~\ref{sec:gaussian} gives an example in which $\nu$ is bounded by a constant as the volume of data, the number of parameters and the number of nodes grow together, with the data split equally among the nodes.

\subsection{Example: Gaussian Data}
\label{sec:gaussian}

We use the following result.

\begin{lemma}[{[\citen{vershynin2012}, Cor.~5.35]}]
\label{lem:vershynin}
Let $\mathbf H\in\mathbb R^{N\times(d+1)}$, $N\ge d+1$, have i.i.d.\ $\mathcal N(0,1)$ entries. For every $t\ge0$, with probability at least $1-2e^{-t^2/2}$,
\[
\sqrt N-\sqrt{d+1}-t\le\sigma_{\min}(\mathbf H)\le\sigma_{\max}(\mathbf H)\le\sqrt N+\sqrt{d+1}+t.
\]
\end{lemma}

Consider a sequence of increasingly large problems, indexed by the number of nodes $L\ge1$. Let $(d_L)_{L\ge1}$ be an increasing sequence of positive integers and $(\varepsilon_L)_{L\ge1}$ a decreasing sequence in $(0,1)$. For each $L\ge1$, let $r_L\ge16(d_L+1)$ and $n_L:=Lr_L$, and let $\mathbf H^{(L)}\in\mathbb R^{n_L\times(d_L+1)}$ have i.i.d.\ $\mathcal N(0,1)$ entries, split equally into $L$ row blocks $\mathbf H^{(L)}_\ell\in\mathbb R^{r_L\times(d_L+1)}$. Set $[\mathbf A^{(L)},\mathbf b^{(L)}]:=\mathbf H^{(L)}/(2\sqrt{n_L})$, so that $\mathbf A^{(L)}$ has $d_L$ columns, and let $\nu_L$ denote the corresponding value of $\nu$.

\begin{proposition}
\label{prop:gaussian}
Let $s_L:=\sqrt{r_L}/4-\sqrt{d_L+1}$, and let $E_L:=\{\normtwo{\mathbf A^{(L)}}>1\}\cup\{\normtwo{\mathbf b^{(L)}}>1\}\cup\{\nu_L>25/9\}$. Then
\begin{equation}
\mathbb P(E_L)\le2(L+1)\,e^{-s_L^2/2}. \label{eq:nu-tail}
\end{equation}
\end{proposition}
\begin{proof}
Apply Lemma~\ref{lem:vershynin} with $t=s_L$ to $\mathbf H^{(L)}$ and to each of the $L$ blocks. By a union bound, all $L+1$ bounds hold outside an event of probability at most $2(L+1)e^{-s_L^2/2}$. Since $\sqrt{d_L+1}+s_L=\sqrt{r_L}/4$ and $\sqrt{n_L}=\sqrt L\sqrt{r_L}$, on this event
\begin{align*}
\sqrt{r_L}\big(\sqrt L-\tfrac14\big)\le\normtwo{\mathbf H^{(L)}}&\le\tfrac54\sqrt{n_L},\\
\normtwo{\mathbf H^{(L)}_\ell}&\le\tfrac54\sqrt{r_L},\qquad\ell=1,\dots,L.
\end{align*}
The upper bound on $\normtwo{\mathbf H^{(L)}}$ gives $\normtwo{[\mathbf A^{(L)},\mathbf b^{(L)}]}\le1$, and hence $\normtwo{\mathbf A^{(L)}}\le1$ and $\normtwo{\mathbf b^{(L)}}\le1$. The remaining two bounds give
\[
\nu_L=\frac{\sum_{\ell=1}^L\normtwo{\mathbf H^{(L)}_\ell}^2}{\normtwo{\mathbf H^{(L)}}^2}\le\frac{25}{16}\cdot\frac{L}{(\sqrt L-\tfrac14)^2}\le\frac{25}{9}. \qedhere
\]
\end{proof}

By~\eqref{eq:nu-tail}, $\mathbb P(E_L)\le\delta'$ whenever $r_L\ge16\big(\sqrt{d_L+1}+\sqrt{2\log(2(L+1)/\delta')}\big)^2$, in particular whenever each node holds a number of rows of order at least $d_L+\log(L/\delta')$.

For each $L$, let $\mathbf S^{(L)}=\blkdiag(\mathbf S^{(L)}_1,\dots,\mathbf S^{(L)}_L)$ be the block-diagonal Gaussian sketch of Theorem~\ref{thm:main} with $\varepsilon=\varepsilon_L$, drawn independently of $\mathbf H^{(L)}$. Outside $E_L$, the hypotheses of Theorem~\ref{thm:main} hold and $\nu_L\le25/9$. Conditionally on $\mathbf H^{(L)}$, the solution $\widetilde{\mathbf x}$ therefore satisfies $F(\widetilde{\mathbf x})\le(1+O(\varepsilon_L))\OPT$ with probability at least $1-\delta$, with total sketch size
\begin{equation}
m=O\big(m_1(\varepsilon_L,\lambda,\delta/2)+m_2(\lambda,\delta/2)\big),\label{eq:gauss-rows}
\end{equation}
which is of the same order as the centralized sketch size $m_{\mathrm c,L}=O(\log(d_L/\delta)/(\varepsilon_L^2\lambda^2))$ for $d_L$ parameters and error $\varepsilon_L$, up to the factor $\log(2/(\varepsilon_L\lambda))$ inside the logarithm.

The communication cost is $m(d_L+1)$ real numbers, up to lower-order terms, which is a fraction of order $1/L$ of the communication $L\,m_{\mathrm c,L}(d_L+1)$ of the baseline. The next result shows that $\nu_L\le25/9$, and hence the guarantee above, holds for every $L$ beyond a random index $L_0$ that is finite almost surely.

\begin{corollary}
\label{cor:asymptotic}
Suppose that
\[
r_L\ge16\big(\sqrt{d_L+1}+\sqrt{6\log(L+1)}\big)^2\qquad\text{for every }L\ge1.
\]
Then there exists a random variable $L_0$, which is a function of $(\mathbf H^{(L)})_{L\ge1}$ and is finite almost surely, such that, for every $L\ge L_0$, $\normtwo{\mathbf A^{(L)}}\le1$, $\normtwo{\mathbf b^{(L)}}\le1$ and $\nu_L\le25/9$. For every such $L$, conditionally on $\mathbf H^{(L)}$, the solution $\widetilde{\mathbf x}$ satisfies $F(\widetilde{\mathbf x})\le(1+O(\varepsilon_L))\OPT$ with probability at least $1-\delta$, with total sketch size~\eqref{eq:gauss-rows}.
\end{corollary}
\begin{proof}
By the choice of $r_L$, $s_L^2/2\ge3\log(L+1)$, so~\eqref{eq:nu-tail} gives $\mathbb P(E_L)\le2(L+1)^{-2}$. The right side is summable in $L$, and by the Borel--Cantelli lemma, $\mathbb P(E_L\ \text{i.o.})=0$, where i.o.\ stands for infinitely often. Let
\[
L_0:=1+\sup_{L\ge1}L\,\mathbf 1_{E_L}.
\]
Since almost surely only finitely many of the indicators $\mathbf 1_{E_L}$ are nonzero, $L_0$ is finite almost surely, and $\mathbf 1_{E_L}=0$ for every $L\ge L_0$. The second claim follows from Theorem~\ref{thm:main}.
\end{proof}

The sketch size~\eqref{eq:gauss-rows} has the same dependence on $d_L$ and $\delta$ as the sketch size of the centralized sketch in Theorem~\ref{thm:mai12}. With $r_L$ at the minimum allowed by Corollary~\ref{cor:asymptotic}, each node holds only $O(d_L+\log L)$ rows. A large volume of data can therefore be sketched with communication of the order of the size of a single centralized sketch, without being stored at a single node.

\subsection{Computation Cost}
\label{sec:cost}

We now compare the computation cost of the two-stage sketch of Theorem~\ref{thm:twostage}, with $m_\ell$ chosen as in Algorithm~\ref{alg:twostage}, with that of the baseline, the distributed implementation of the centralized sketch of Section~\ref{sec:baseline}.
We suppose that $\widetilde m\le C_2\,\mc$ for an absolute constant $C_2$, that is, the sketch size of the second stage is at most a constant multiple of that of the centralized sketch. This holds because both are of order $\log(d/\delta)/(\varepsilon^2\lambda^2)$ (Theorems~\ref{thm:mai12} and~\ref{thm:twostage}).

\subsubsection{Cost of the Baseline}
Node $\ell$ forms $\mathbf G_\ell[\mathbf A_\ell,\mathbf b_\ell]$, where $\mathbf G_\ell\in\mathbb R^{\mc\times n_\ell}$, at a cost of $2(d+1)\,\mc\,n_\ell$ flops. Summing over $\ell$, the cost of the baseline is at least $C_{\mathrm{baseline}}:=2(d+1)\,\mc\,n$ flops.

\subsubsection{Cost of the Two-Stage Sketch}
The two-stage sketch requires
\[
\underbrace{2(d+1)\sum_{\ell=1}^L n_\ell\,m_\ell}_{(a)}+\underbrace{2(d+1)\,\widetilde m\,m}_{(b)}+\underbrace{O\big(nd\log(d+1)\log(L/\delta)\big)}_{(c)}
\]
flops, apart from $2n$ flops for the scalars $\normtwo{\mathbf b_\ell}^2$. Term~(a) is the cost of forming $\widehat{\mathbf S}_\ell[\mathbf A_\ell,\mathbf b_\ell]$ at the nodes. Term~(b) is the cost of forming $\widetilde{\mathbf S}(\widehat{\mathbf S}[\mathbf A,\mathbf b])$ at the central node. Term~(c) is the cost of estimating $\eta_1,\dots,\eta_L$. An estimate within a factor of $2$ suffices, and block Krylov iteration gives such an estimate with probability at least $1-\delta$ at this cost (Section~\ref{sec:etaest} and Lemma~\ref{lem:etaest}).

\subsubsection{Bound on the Total Sketch Size of the First Stage}
Since $\normtwo{\mathbf A}\le1$ and $\normtwo{\mathbf b}\le1$, for every $\mathbf y=(\mathbf x,y_{d+1})\in\mathbb R^{d+1}$,
\[
\normtwo{[\mathbf A,\mathbf b]\mathbf y}\le\normtwo{\mathbf x}+|y_{d+1}|\le\sqrt2\,\normtwo{\mathbf y},
\]
so $\normtwo{[\mathbf A,\mathbf b]}^2\le2$ and $\eta_\Sigma=\nu\normtwo{[\mathbf A,\mathbf b]}^2/4\le\nu/2$. Comparing~\eqref{eq:mafun}--\eqref{eq:mbfun} with~\eqref{eq:mtildea}--\eqref{eq:mtildeb}, there is an absolute constant $C_1$ such that $m_1(\varepsilon/2,\lambda,\delta/4)\le C_1\,\widetilde m$ and $m_2(3\lambda/7,\delta/4)\le C_1\,\widetilde m$. Bounding the maximum in $m_\ell$ by the sum of its two terms gives $m_\ell\le C_1\,\widetilde m\,(\eta_\ell+\beta_\ell)$. Summing over $\ell$, and using $\sum_\ell\beta_\ell\le1$ and $\nu\ge1$,
\[
m\le C_1\,\widetilde m\,(\eta_\Sigma+1)\le C_1\,\widetilde m\,\big(\tfrac{\nu}{2}+1\big)\le\tfrac32\,C_1\,\widetilde m\,\nu.
\]
If $m_\ell$ is computed from estimates of $\eta_\ell$ within a factor of $2$, the same bound holds with $C_1$ replaced by $2C_1$.

\subsubsection{Comparison of the Terms}
Suppose that the rows are split equally, with $n_\ell=n/L$ for every $\ell$. Then Term~(a) equals $2(d+1)\,m\,n/L$. By the bound on $m$ and $\widetilde m\le C_2\,\mc$, it is at most $\tfrac32C_1C_2\,\nu/L$ times $C_{\mathrm{baseline}}$. It is therefore much smaller than $C_{\mathrm{baseline}}$ when $\nu\ll L$.

Term~(b) is at most $3C_1(d+1)\,\widetilde m^2\nu\le3C_1C_2(d+1)\,\widetilde m\,\mc\,\nu$, which is $\tfrac32C_1C_2\,\nu\,\widetilde m/n$ times $C_{\mathrm{baseline}}$. If each node holds at least $\widetilde m$ rows, that is, $n/L\ge\widetilde m$, this ratio is at most $\tfrac32C_1C_2\,\nu/L$, so Term~(b) is also much smaller than $C_{\mathrm{baseline}}$ when $\nu\ll L$.

For Term~(c), $\mc\ge\widetilde m/C_2$ and~\eqref{eq:mtildea} give $\mc\ge C_3\log(2d+2)/(\varepsilon^2\lambda^2)$ for an absolute constant $C_3>0$. Since $\log(d+1)\le\log(2d+2)$, Term~(c) is at most $O\big(\varepsilon^2\lambda^2\log(L/\delta)\big)$ times $C_{\mathrm{baseline}}$, so it is much smaller than $C_{\mathrm{baseline}}$ when $\varepsilon^2\lambda^2\log(L/\delta)\ll1$. The $2n$ flops for the scalars are $1/((d+1)\mc)$ times $C_{\mathrm{baseline}}$, which is negligible.

Hence, when the heterogeneity measure is much smaller than the number of nodes ($\nu\ll L$), each node holds many rows ($n/L\ge\widetilde m$), and the required accuracy is high ($\varepsilon^2\lambda^2\log(L/\delta)\ll1$), the two-stage sketch requires less computation than the baseline, and Terms~(a) and~(b) are smaller than $C_{\mathrm{baseline}}$ by a factor of order $L/\nu$.
\section{Proofs of the Main Results}
\label{sec:partI}

This section proves the results of Section~\ref{sec:mainresult}. Section~\ref{sec:proofoutline} outlines the proofs. Section~\ref{sec:etaest} then shows how each node estimates $\eta_\ell$.

\subsection{Proof Outline}
\label{sec:proofoutline}

The proofs have six steps.
\begin{enumerate}
\item Section~\ref{sec:subgrad} provides intermediate results that hold for every $\mathbf S$ (Lemma~\ref{lem:apriori} and Proposition~\ref{prop:errdecomp}).
\item Section~\ref{sec:partI-distortion} proves that C1$(\varepsilon,\lambda)$ and C2$(\lambda)$ together imply $F(\widetilde{\mathbf x})\le(1+O(\varepsilon))\,\OPT$ (Theorem~\ref{thm:partI}).
\item Section~\ref{sec:partI-chaining} proves Theorem~\ref{thm:mamb}: the block-diagonal Gaussian sketch satisfies C1$(\varepsilon,\lambda)$ (part~(i)) and C2$(\lambda)$ (part~(ii)), each with probability at least $1-\delta$.
\item Section~\ref{sec:mainproof} combines Steps~2 and~3 to prove Theorem~\ref{thm:main}.
\item Section~\ref{sec:twostage} proves Theorem~\ref{thm:transfer}: if $\widehat{\mathbf S}$ satisfies C1$(\varepsilon/2,\lambda)$ and C2$(3\lambda/7)$, then $\widetilde{\mathbf S}\widehat{\mathbf S}$ satisfies C1$(\varepsilon,\lambda)$ and C2$(\lambda)$.
\item Section~\ref{sec:twostageproof} combines Step~3 with Theorems~\ref{thm:transfer} and~\ref{thm:partI} to prove Theorem~\ref{thm:twostage}.
\end{enumerate}

\subsection{Intermediate Results}
\label{sec:subgrad}

The results in this subsection hold for every $\mathbf S$. They are deterministic and do not use C1 or C2.

\begin{lemma}
\label{lem:apriori}
Under the assumptions of Section~\ref{sec:notation}, the following hold for every $\mathbf S$.

\titem{i}{lem:feasxstar}$\normtwo{\mathbf A\mathbf x^*-\mathbf b}\le1$ and $\normtwo{\mathbf A\mathbf x^*}\le2$.
\titem{ii}{lem:feasxtilde}
\[
\normtwo{\mathbf S\mathbf A\widetilde{\mathbf x}-\mathbf S\mathbf b}\le\normtwo{\mathbf S\mathbf b},\qquad\normtwo{\mathbf S\mathbf A\widetilde{\mathbf x}}\le2\normtwo{\mathbf S\mathbf b}.
\]
\titem{iii}{lem:bnorm}$\normtwo{\mathbf b}^2\le\OPT+2\|\mathbf x^*\|_1\le3\,\OPT/\lambda$.
\titem{iv}{lem:optle1}$\OPT\le1$.
\end{lemma}
\begin{proof}
\emph{(i)} Since $\mathbf x^*$ minimizes $F$, $\normtwo{\mathbf A\mathbf x^*-\mathbf b}^2\le F(\mathbf x^*)\le F(\mathbf 0)=\normtwo{\mathbf b}^2\le1$. The triangle inequality then gives $\normtwo{\mathbf A\mathbf x^*}\le\normtwo{\mathbf A\mathbf x^*-\mathbf b}+\normtwo{\mathbf b}\le2$.

\emph{(ii)} The argument for~(i) applies to the sketched objective, whose minimizer is $\widetilde{\mathbf x}$ and whose value at $\mathbf x=\mathbf 0$ is $\normtwo{\mathbf S\mathbf b}^2$.

\emph{(iii)} We have
\[
\normtwo{\mathbf b}^2\lestep{a}\OPT+2\langle\mathbf A\mathbf x^*,\mathbf b\rangle\lestep{b}\OPT+2\|\mathbf x^*\|_1\lestep{c}\frac{3\,\OPT}{\lambda}.
\]
Step~(a) follows from $\normtwo{\mathbf A\mathbf x^*-\mathbf b}^2\le\OPT$. Step~(b) follows from $\langle\mathbf A\mathbf x^*,\mathbf b\rangle=\sum_ix_i^*\langle\mathbf a_i,\mathbf b\rangle$ and H\"older's inequality, with $|\langle\mathbf a_i,\mathbf b\rangle|\le\normtwo{\mathbf a_i}\normtwo{\mathbf b}\le1$. Step~(c) holds because $\lambda\|\mathbf x^*\|_1\le\OPT$ and $\lambda<1$.

\emph{(iv)} Since $\mathbf x^*$ minimizes $F$, $\OPT=F(\mathbf x^*)\le F(\mathbf 0)=\normtwo{\mathbf b}^2\le1$.
\end{proof}

\begin{proposition}
\label{prop:errdecomp}
Let $\mathbf A$, $\mathbf b$ and $\lambda$ be as in Section~\ref{sec:notation}, let $\mathbf S\in\mathbb R^{m\times n}$ be arbitrary, let $\mathbf u:=\widetilde{\mathbf x}-\mathbf x^*$, and let
\begin{align*}
\alpha&:=\normtwo{\mathbf A\widetilde{\mathbf x}}^2-\normtwo{\mathbf S\mathbf A\widetilde{\mathbf x}}^2,\\
\rho^+&:=\normtwo{\mathbf A(\widetilde{\mathbf x}+\mathbf x^*)}^2-\normtwo{\mathbf S\mathbf A(\widetilde{\mathbf x}+\mathbf x^*)}^2,\\
\rho^-&:=\normtwo{\mathbf A(\widetilde{\mathbf x}-\mathbf x^*)}^2-\normtwo{\mathbf S\mathbf A(\widetilde{\mathbf x}-\mathbf x^*)}^2.
\end{align*}
Then
\begin{tcmultline}
F(\widetilde{\mathbf x})\le\OPT+2\Big(|\alpha|+\tfrac14|\rho^-|+\tfrac14|\rho^+| \tcbr
+\max_{1\le i\le d}\big|\langle\mathbf a_i,\mathbf b\rangle-\langle\mathbf S\mathbf a_i,\mathbf S\mathbf b\rangle\big|\,\|\widetilde{\mathbf x}-\mathbf x^*\|_1\Big). \label{eq:errdecomp}
\end{tcmultline}
\end{proposition}
\begin{proof}
Let $f(\mathbf x):=\normtwo{\mathbf A\mathbf x-\mathbf b}^2$ and
\[
\Delta:=\big\langle \mathbf A^\top(\mathbf A\widetilde{\mathbf x}-\mathbf b),\mathbf u\big\rangle-\big\langle \mathbf A^\top\mathbf S^\top(\mathbf S\mathbf A\widetilde{\mathbf x}-\mathbf S\mathbf b),\mathbf u\big\rangle.
\]
We first show that $F(\widetilde{\mathbf x})\le\OPT+2\Delta$.
Since $f$ is convex,
\begin{equation}
f(\widetilde{\mathbf x})\le f(\mathbf x^*)+2\big\langle\mathbf A^\top(\mathbf A\widetilde{\mathbf x}-\mathbf b),\mathbf u\big\rangle. \label{eq:subgrad-i}
\end{equation}
By the optimality of $\widetilde{\mathbf x}$ for the sketched objective, $\lambda\mathbf g=-2\mathbf A^\top\mathbf S^\top(\mathbf S\mathbf A\widetilde{\mathbf x}-\mathbf S\mathbf b)$ with $\mathbf g\in\partial\|\widetilde{\mathbf x}\|_1$. By convexity of $\|\cdot\|_1$,
\begin{equation}
\lambda\|\widetilde{\mathbf x}\|_1\le\lambda\|\mathbf x^*\|_1+\lambda\langle \mathbf g,\mathbf u\rangle, \label{eq:subgrad-ii}
\end{equation}
and, by the choice of $\mathbf g$,
\begin{equation}
\lambda\langle\mathbf g,\mathbf u\rangle=-2\big\langle\mathbf A^\top\mathbf S^\top(\mathbf S\mathbf A\widetilde{\mathbf x}-\mathbf S\mathbf b),\mathbf u\big\rangle. \label{eq:subgrad-iii}
\end{equation}
We add \eqref{eq:subgrad-i} and \eqref{eq:subgrad-ii} and substitute \eqref{eq:subgrad-iii} to obtain $F(\widetilde{\mathbf x})\le\OPT+2\Delta$.

Next, we write $\Delta$ using $\alpha$, $\rho^+$ and $\rho^-$.
\begin{tcmultline*}
\Delta=\big\langle\mathbf A^\top(\mathbf I-\mathbf S^\top\mathbf S)\mathbf A\widetilde{\mathbf x},\mathbf u\big\rangle
+\big\langle\mathbf A^\top\mathbf b-\mathbf A^\top\mathbf S^\top\mathbf S\mathbf b,\mathbf x^*-\widetilde{\mathbf x}\big\rangle.
\end{tcmultline*}
With $\mathbf u=\widetilde{\mathbf x}-\mathbf x^*$, the first inner product expands as
\begin{tcmultline*}
\big\langle\mathbf A^\top(\mathbf I-\mathbf S^\top\mathbf S)\mathbf A\widetilde{\mathbf x},\mathbf u\big\rangle=\alpha \tcbr
-\big(\langle\mathbf A\widetilde{\mathbf x},\mathbf A\mathbf x^*\rangle \tcbr
-\langle\mathbf S\mathbf A\widetilde{\mathbf x},\mathbf S\mathbf A\mathbf x^*\rangle\big).
\end{tcmultline*}
The identity $4\langle\mathbf p,\mathbf q\rangle=\normtwo{\mathbf p+\mathbf q}^2-\normtwo{\mathbf p-\mathbf q}^2$, applied to $\langle\mathbf A\widetilde{\mathbf x},\mathbf A\mathbf x^*\rangle$ and $\langle\mathbf S\mathbf A\widetilde{\mathbf x},\mathbf S\mathbf A\mathbf x^* \rangle$, yields
\[
\Delta=\alpha-\tfrac14\rho^++\tfrac14\rho^-+\big\langle\mathbf A^\top\mathbf b-\mathbf A^\top\mathbf S^\top\mathbf S\mathbf b,\mathbf x^*-\widetilde{\mathbf x}\big\rangle.
\]

Finally, the second inner product equals $\sum_{i=1}^d(x_i^*-\widetilde x_i)(\langle\mathbf a_i,\mathbf b\rangle-\langle\mathbf S\mathbf a_i,\mathbf S\mathbf b\rangle)$, and H\"older's inequality bounds its absolute value by the last term in~\eqref{eq:errdecomp}. Bounding the remaining terms of $\Delta$ by their absolute values yields an upper bound on $\Delta$, and substituting it into $F(\widetilde{\mathbf x})\le\OPT+2\Delta$ gives~\eqref{eq:errdecomp}.
\end{proof}

\subsection{Sufficiency of C1 and C2}
\label{sec:partI-distortion}

This subsection proves Theorem~\ref{thm:partI}. Each of the four terms of~\eqref{eq:errdecomp} is at most an absolute constant multiple of $\varepsilon\,\OPT$.

The next two lemmas collect consequences of C1$(\varepsilon,\lambda)$ and C2$(\lambda)$.

\begin{lemma}
\label{lem:conseqC1}
Let $\mathbf A$, $\mathbf b$ and $\lambda$ be as in Section~\ref{sec:notation}, and suppose that $\mathbf S$ satisfies C1$(\varepsilon,\lambda)$. Then the following hold.

\titem{i}{lem:Sb}$\normtwo{\mathbf S\mathbf b}\le\normtwo{\mathbf b}+\varepsilon\lambda\le1+\varepsilon\lambda$.
\titem{ii}{lem:homext}For every $\mathbf v\in\mathbb R^d$, $\big|\normtwo{\mathbf S\mathbf A\mathbf v}-\normtwo{\mathbf A\mathbf v}\big|\le\varepsilon\lambda\|\mathbf v\|_1$.
\titem{iii}{lem:pairwise}For every $1\le i\le d$, $\big|\normtwo{\mathbf S\mathbf a_i\mp\mathbf S\mathbf b}-\normtwo{\mathbf a_i\mp\mathbf b}\big|\le2\varepsilon\lambda$.
\titem{iv}{lem:cross}For every $1\le i\le d$, $\big|\langle\mathbf a_i,\mathbf b\rangle-\langle\mathbf S\mathbf a_i,\mathbf S\mathbf b\rangle\big|\le6\varepsilon\lambda$. For every $\mathbf v\in\mathbb R^d$, $\big|\langle\mathbf A\mathbf v,\mathbf b\rangle-\langle\mathbf S\mathbf A\mathbf v,\mathbf S\mathbf b\rangle\big|\le6\varepsilon\lambda\|\mathbf v\|_1$.
\end{lemma}
\begin{proof}
\emph{(i)} Since $\mathbf b\in T$, C1$(\varepsilon,\lambda)$ at $\mathbf z=\mathbf b$ gives $\normtwo{\mathbf S\mathbf b}\le\normtwo{\mathbf b}+\varepsilon\lambda\le1+\varepsilon\lambda$.

\emph{(ii)} If $\mathbf v=\mathbf 0$, the inequality is trivial. Otherwise, $\mathbf A\mathbf v/\|\mathbf v\|_1$ is a convex combination of $\pm\mathbf a_1,\dots,\pm\mathbf a_d$, so it lies in $T$. Applying C1$(\varepsilon,\lambda)$ to this point gives the claim.

\emph{(iii)} The point $\tfrac12(\mathbf a_i\mp\mathbf b)$ lies in $T$. Applying C1$(\varepsilon,\lambda)$ to it gives the claim.

\emph{(iv)} Fix $i$. Since $\normtwo{\mathbf a_i\mp\mathbf b}\le2$ by Section~\ref{sec:notation} and $\normtwo{\mathbf S\mathbf a_i\mp\mathbf S\mathbf b}\le2+2\varepsilon\lambda$ by part~(iii), we have
\begin{tcmultline*}
\big|\normtwo{\mathbf a_i\mp\mathbf b}^2-\normtwo{\mathbf S\mathbf a_i\mp\mathbf S\mathbf b}^2\big| \tcbr
\lestep{a}(2\varepsilon\lambda)(4+2\varepsilon\lambda)\lestep{b}12\varepsilon\lambda,
\end{tcmultline*}
where (a) follows from $s^2-t^2=(s-t)(s+t)$ and part~(iii), and (b) holds since $\varepsilon\lambda<1$. For any $\mathbf p,\mathbf q$, $4\langle\mathbf p,\mathbf q\rangle=\normtwo{\mathbf p+\mathbf q}^2-\normtwo{\mathbf p-\mathbf q}^2$. Using this identity for $\langle\mathbf a_i,\mathbf b\rangle$ and for $\langle\mathbf S\mathbf a_i,\mathbf S\mathbf b\rangle$, we obtain
\begin{tcmultline*}
4\big(\langle\mathbf a_i,\mathbf b\rangle-\langle\mathbf S\mathbf a_i,\mathbf S\mathbf b\rangle\big)=\big[\normtwo{\mathbf a_i+\mathbf b}^2-\normtwo{\mathbf S\mathbf a_i+\mathbf S\mathbf b}^2\big] \tcbr
-\big[\normtwo{\mathbf a_i-\mathbf b}^2-\normtwo{\mathbf S\mathbf a_i-\mathbf S\mathbf b}^2\big].
\end{tcmultline*}
Both brackets are bounded by $12\varepsilon\lambda$ in absolute value, and the first claim follows. For the second claim, $\langle\mathbf A\mathbf v,\mathbf b\rangle-\langle\mathbf S\mathbf A\mathbf v,\mathbf S\mathbf b\rangle=\sum_iv_i(\langle\mathbf a_i,\mathbf b\rangle-\langle\mathbf S\mathbf a_i,\mathbf S\mathbf b\rangle)$, so H\"older's inequality and the first claim give the bound $6\varepsilon\lambda\|\mathbf v\|_1$.
\end{proof}

\begin{lemma}
\label{lem:l1ctrl}
Let $\mathbf A$, $\mathbf b$ and $\lambda$ be as in Section~\ref{sec:notation}, and suppose that $\mathbf S$ satisfies C1$(\varepsilon,\lambda)$ and C2$(\lambda)$. Then the following hold.

\titem{i}{lem:sketchcompI}
\[
\normtwo{\mathbf S(\mathbf A\mathbf x^*-\mathbf b)}^2+\lambda\|\mathbf x^*\|_1\le(2+17\varepsilon)\,\OPT.
\]
\titem{ii}{lem:l1boundsI}
\[
\|\mathbf x^*\|_1\le\frac{\OPT}{\lambda},\qquad\|\widetilde{\mathbf x}\|_1\le(2+17\varepsilon)\frac{\OPT}{\lambda},
\]
\[
\|\widetilde{\mathbf x}-\mathbf x^*\|_1\le(3+17\varepsilon)\frac{\OPT}{\lambda},\qquad\|\widetilde{\mathbf x}+\mathbf x^*\|_1\le(3+17\varepsilon)\frac{\OPT}{\lambda}.
\]
\end{lemma}
\begin{proof}
\emph{(i)} Adding and subtracting $\normtwo{\mathbf A\mathbf x^*-\mathbf b}^2$ gives
\begin{multline*}
\normtwo{\mathbf S(\mathbf A\mathbf x^*-\mathbf b)}^2+\lambda\|\mathbf x^*\|_1=\OPT \tcbr
+\underbrace{\big(\normtwo{\mathbf S\mathbf A\mathbf x^*}^2-\normtwo{\mathbf A\mathbf x^*}^2\big)}_{R_1} \\
-2\underbrace{\big(\langle\mathbf S\mathbf A\mathbf x^*,\mathbf S\mathbf b\rangle-\langle\mathbf A\mathbf x^*,\mathbf b\rangle\big)}_{R_2} \tcbr
+\underbrace{\big(\normtwo{\mathbf S\mathbf b}^2-\normtwo{\mathbf b}^2\big)}_{R_3}.
\end{multline*}
\emph{Bounding $R_1$.} Lemma~\ref{lem:homext} with $\mathbf v=\mathbf x^*$, together with $\lambda\|\mathbf x^*\|_1\le\OPT$, gives $|\normtwo{\mathbf S\mathbf A\mathbf x^*}-\normtwo{\mathbf A\mathbf x^*}|\le\varepsilon\,\OPT$. By Lemma~\ref{lem:feasxstar}, $\normtwo{\mathbf A\mathbf x^*}\le2$. Since $\varepsilon\,\OPT\le1$, it follows that $\normtwo{\mathbf S\mathbf A\mathbf x^*}+\normtwo{\mathbf A\mathbf x^*}\le4+\varepsilon\,\OPT\le5$. By $s^2-t^2=(s-t)(s+t)$, $|R_1|\le5\varepsilon\,\OPT$.

\emph{Bounding $R_2$.} Lemma~\ref{lem:cross} with $\mathbf v=\mathbf x^*$ gives $|R_2|\le6\varepsilon\lambda\|\mathbf x^*\|_1\le6\varepsilon\,\OPT$. Hence $2|R_2|\le12\varepsilon\,\OPT$.

\emph{Bounding $R_3$.} C2$(\lambda)$ and Lemma~\ref{lem:bnorm} give $|R_3|\le\tfrac{\lambda}{3}\normtwo{\mathbf b}^2\le\tfrac{\lambda}{3}\cdot3\,\OPT/\lambda=\OPT$.

The left side is at most $\OPT+|R_1|+2|R_2|+|R_3|\le\OPT+5\varepsilon\,\OPT+12\varepsilon\,\OPT+\OPT=(2+17\varepsilon)\,\OPT$.

\emph{(ii)} The inequality $\lambda\|\mathbf x^*\|_1\le\OPT$ holds because both terms in $F(\mathbf x^*)$ are nonnegative. For $\widetilde{\mathbf x}$,
\begin{tcmultline*}
\lambda\|\widetilde{\mathbf x}\|_1\le\normtwo{\mathbf S(\mathbf A\widetilde{\mathbf x}-\mathbf b)}^2+\lambda\|\widetilde{\mathbf x}\|_1 \tcbr
\lestep{a}\normtwo{\mathbf S(\mathbf A\mathbf x^*-\mathbf b)}^2+\lambda\|\mathbf x^*\|_1\lestep{b}(2+17\varepsilon)\,\OPT.
\end{tcmultline*}
Step~(a) uses the optimality of $\widetilde{\mathbf x}$ for the sketched objective. Step~(b) is part~(i). The last two inequalities follow from the first two and the triangle inequality for $\|\cdot\|_1$.
\end{proof}

The next proposition bounds each of the four terms on the right side of~\eqref{eq:errdecomp} in terms of $\varepsilon$ and $\OPT$.

\begin{proposition}
\label{prop:fourterms}
Let $\mathbf A$, $\mathbf b$ and $\lambda$ be as in Section~\ref{sec:notation}, and suppose that $\mathbf S$ satisfies C1$(\varepsilon,\lambda)$ and C2$(\lambda)$. Then the following hold.

\titem{a}{prop:terma}$|\alpha|\le(2+17\varepsilon)(10+17\varepsilon)\,\varepsilon\,\OPT$.
\titem{b}{prop:termb}$|\rho^-|\le(3+17\varepsilon)(15+17\varepsilon)\,\varepsilon\,\OPT$.
\titem{c}{prop:termc}$|\rho^+|\le(3+17\varepsilon)(15+17\varepsilon)\,\varepsilon\,\OPT$.
\titem{d}{prop:termd}
\begin{tcmultline*}
\max_{1\le i\le d}\big|\langle\mathbf a_i,\mathbf b\rangle-\langle\mathbf S\mathbf a_i,\mathbf S\mathbf b\rangle\big|\cdot\|\widetilde{\mathbf x}-\mathbf x^*\|_1 \tcbr
\le6(3+17\varepsilon)\,\varepsilon\,\OPT.
\end{tcmultline*}
\end{proposition}
\begin{proof}
\emph{(a)} By Lemma~\ref{lem:homext} with $\mathbf v=\widetilde{\mathbf x}$ and Lemma~\ref{lem:l1boundsI},
\begin{equation}
\big|\normtwo{\mathbf S\mathbf A\widetilde{\mathbf x}}-\normtwo{\mathbf A\widetilde{\mathbf x}}\big|\le\varepsilon\lambda\|\widetilde{\mathbf x}\|_1\le(2+17\varepsilon)\,\varepsilon\,\OPT. \label{eq:terma-diff}
\end{equation}
Next, we bound the sum of the two norms:
\begin{multline}
\normtwo{\mathbf S\mathbf A\widetilde{\mathbf x}}+\normtwo{\mathbf A\widetilde{\mathbf x}}\lestep{i}2\normtwo{\mathbf S\mathbf A\widetilde{\mathbf x}}+(2+17\varepsilon)\,\varepsilon\,\OPT \tcbr
\lestep{ii}4\normtwo{\mathbf S\mathbf b}+(2+17\varepsilon)\,\varepsilon\,\OPT \\
\lestep{iii}4\normtwo{\mathbf b}+4\varepsilon\lambda+(2+17\varepsilon)\,\varepsilon\,\OPT\lestep{iv}10+17\varepsilon. \label{eq:terma-sum}
\end{multline}
Step~(i) uses \eqref{eq:terma-diff}. Step~(ii) is Lemma~\ref{lem:feasxtilde}. Step~(iii) is Lemma~\ref{lem:Sb}. Step~(iv) uses $\normtwo{\mathbf b}\le1$, $\varepsilon\lambda<1$ and $\varepsilon\,\OPT\le1$. Finally, $|\alpha|$ is a difference of squares. Writing it as $(s-t)(s+t)$ and applying \eqref{eq:terma-diff} and \eqref{eq:terma-sum} gives $|\alpha|\le\big((2+17\varepsilon)\,\varepsilon\,\OPT\big)(10+17\varepsilon)$.

\emph{(b)} By Lemma~\ref{lem:homext} with $\mathbf v=\mathbf u$ and Lemma~\ref{lem:l1boundsI},
\begin{equation}
\big|\normtwo{\mathbf S\mathbf A\mathbf u}-\normtwo{\mathbf A\mathbf u}\big|\le\varepsilon\lambda\|\mathbf u\|_1\le(3+17\varepsilon)\,\varepsilon\,\OPT. \label{eq:termb-diff}
\end{equation}
Splitting $\mathbf u=\widetilde{\mathbf x}-\mathbf x^*$ with the triangle inequality gives
\begin{tcmultline*}
\normtwo{\mathbf S\mathbf A\mathbf u}+\normtwo{\mathbf A\mathbf u}\le\underbrace{\normtwo{\mathbf S\mathbf A\widetilde{\mathbf x}}+\normtwo{\mathbf A\widetilde{\mathbf x}}}_{U_1} \tcbr
+\underbrace{\normtwo{\mathbf S\mathbf A\mathbf x^*}+\normtwo{\mathbf A\mathbf x^*}}_{U_2}.
\end{tcmultline*}
As in \eqref{eq:terma-sum}, $U_1\le2\normtwo{\mathbf S\mathbf A\widetilde{\mathbf x}}+(2+17\varepsilon)\,\varepsilon\,\OPT$. For $U_2$, Lemmas~\ref{lem:homext} (with $\mathbf v=\mathbf x^*$) and~\ref{lem:l1boundsI} give $|\normtwo{\mathbf S\mathbf A\mathbf x^*}-\normtwo{\mathbf A\mathbf x^*}|\le\varepsilon\lambda\|\mathbf x^*\|_1\le\varepsilon\,\OPT$. Hence $U_2\le2\normtwo{\mathbf A\mathbf x^*}+\varepsilon\,\OPT$. Therefore
\begin{multline}
\normtwo{\mathbf S\mathbf A\mathbf u}+\normtwo{\mathbf A\mathbf u}\lestep{i}4\normtwo{\mathbf S\mathbf b}+4\normtwo{\mathbf b}+(3+17\varepsilon)\,\varepsilon\,\OPT \\
\lestep{ii}8\normtwo{\mathbf b}+4\varepsilon\lambda+(3+17\varepsilon)\,\varepsilon\,\OPT\lestep{iii}15+17\varepsilon. \label{eq:termb-sum}
\end{multline}
Step~(i) adds the bounds on $U_1$ and $U_2$. It then uses Lemma~\ref{lem:feasxtilde} for $\normtwo{\mathbf S\mathbf A\widetilde{\mathbf x}}$, and $\normtwo{\mathbf A\mathbf x^*}\le\normtwo{\mathbf A\mathbf x^*-\mathbf b}+\normtwo{\mathbf b}\le2\normtwo{\mathbf b}$ from the proof of Lemma~\ref{lem:feasxstar}. Step~(ii) is Lemma~\ref{lem:Sb}. Step~(iii) uses $\normtwo{\mathbf b}\le1$, $\varepsilon\lambda<1$ and $\varepsilon\,\OPT\le1$. Finally, writing $|\rho^-|$ as $(s-t)(s+t)$ and applying \eqref{eq:termb-diff} and \eqref{eq:termb-sum} gives $|\rho^-|\le\big((3+17\varepsilon)\,\varepsilon\,\OPT\big)(15+17\varepsilon)$.

\emph{(c)} The proof of~(b) applies with $\widetilde{\mathbf x}+\mathbf x^*$ in place of $\widetilde{\mathbf x}-\mathbf x^*$, since Lemma~\ref{lem:l1boundsI} gives $\|\widetilde{\mathbf x}+\mathbf x^*\|_1\le(3+17\varepsilon)\,\OPT/\lambda$.

\emph{(d)} By Lemma~\ref{lem:cross}, the maximum is at most $6\varepsilon\lambda$. By Lemma~\ref{lem:l1boundsI}, $\|\widetilde{\mathbf x}-\mathbf x^*\|_1\le(3+17\varepsilon)\,\OPT/\lambda$. The product of the two bounds gives the claim.
\end{proof}

We now prove Theorem~\ref{thm:partI} by combining Propositions~\ref{prop:errdecomp} and~\ref{prop:fourterms}.

\begin{proof}[Proof of Theorem~\ref{thm:partI}]
Let
\begin{tcmultline*}
q(\varepsilon):=\underbrace{(2+17\varepsilon)(10+17\varepsilon)}_{(a)} \tcbr
+\underbrace{\tfrac12(3+17\varepsilon)(15+17\varepsilon)}_{(b)+(c)}+\underbrace{6(3+17\varepsilon)}_{(d)},
\end{tcmultline*}
where each brace collects the bound of Proposition~\ref{prop:fourterms} on the corresponding term of~\eqref{eq:errdecomp}, divided by $\varepsilon\,\OPT$. Substituting these bounds into~\eqref{eq:errdecomp}, we obtain $F(\widetilde{\mathbf x})\le\OPT+2\,q(\varepsilon)\,\varepsilon\,\OPT$. Expanding,
\begin{tcmultline*}
2\,q(\varepsilon)=(40+408\varepsilon+578\varepsilon^2)+(45+306\varepsilon+289\varepsilon^2) \tcbr
+(36+204\varepsilon)=121+918\varepsilon+867\varepsilon^2,
\end{tcmultline*}
which gives the first inequality. The second holds since $121+918\varepsilon+867\varepsilon^2\le1906$ for $\varepsilon\in(0,1)$.
\end{proof}

\subsection{Proof of Theorem~\ref{thm:mamb}}
\label{sec:partI-chaining}

This subsection proves Theorem~\ref{thm:mamb}. Part~(i) is proved using the tail bound of~\cite[Thm.~6.5]{dirksen2015}. Part~(ii) concerns the single vector $\mathbf b$ and follows from Lemma~\ref{lem:srinivasa}.

\subsubsection{Proof of Theorem~\ref{thm:mamb}(i)}
\label{sec:proof-ma}
\label{sec:tools}

\emph{Preliminaries.} We use the sub-Gaussian norm of Section~\ref{subsec:notation-conventions}. For a set $\mathcal A$ of $p\times q$ matrices, let $\mathsf d_\infty(\mathbf M_1,\mathbf M_2):=\normtwo{\mathbf M_1-\mathbf M_2}$ and
\begin{align*}
\bar\Delta_2(\mathcal A)&:=\sup_{\mathbf M\in\mathcal A}\normF{\mathbf M},\\
\bar\Delta_4(\mathcal A)&:=\sup_{\mathbf M\in\mathcal A}\normSp{\mathbf M}{4},\\
\bar\Delta_\infty(\mathcal A)&:=\sup_{\mathbf M\in\mathcal A}\normtwo{\mathbf M}.
\end{align*}
For a metric $\mathsf d$ on a set $\mathcal T$, a point $\mathbf x\in\mathcal T$ and a nonempty finite set $\mathcal S\subset\mathcal T$, the distance from $\mathbf x$ to $\mathcal S$ is
\begin{equation}
\mathsf d(\mathbf x,\mathcal S):=\min_{\mathbf s\in\mathcal S}\mathsf d(\mathbf x,\mathbf s). \label{eq:setdist}
\end{equation}
The $\gamma_2$-functional~\cite{talagrand2021} is
\[
\gamma_2(\mathcal T,\mathsf d):=\inf_{\{\mathcal S_r\}_{r=0}^\infty}\sup_{\mathbf x\in\mathcal T}\sum_{r\ge0}2^{r/2}\mathsf d(\mathbf x,\mathcal S_r),
\]
where the infimum is over admissible sequences, that is, subsets $\mathcal S_r\subset\mathcal T$, $r\ge0$, with $|\mathcal S_0|=1$ and $|\mathcal S_r|\le2^{2^r}$~\cite{dirksen2015}.

\begin{theorem}[{[\citen{dirksen2015}, Thm.~6.5]}]
\label{thm:dirksen}
There exist absolute constants $c_7,c_8>0$ such that the following holds. Let $\mathcal A$ be a set of $p\times q$ matrices, and let $\boldsymbol\xi=(\xi_1,\dots,\xi_q)$ have independent, real-valued, mean-zero entries. Then, for every $h\ge1$,
\begin{multline*}
\mathbb P\bigg(\sup_{\mathbf M\in\mathcal A}\big|\normtwo{\mathbf M\boldsymbol\xi}^2-\mathbb E\normtwo{\mathbf M\boldsymbol\xi}^2\big|\ge \\
c_7\|\boldsymbol\xi\|_{\psi_2}^2\Big(\gamma_2^2(\mathcal A,\mathsf d_\infty)+\bar\Delta_2(\mathcal A)\gamma_2(\mathcal A,\mathsf d_\infty)\Big) \tcbr
+c_8\|\boldsymbol\xi\|_{\psi_2}^2\Big(\sqrt{h}\,\bar\Delta_4^2(\mathcal A)+h\,\bar\Delta_\infty^2(\mathcal A)\Big)\bigg)\le e^{-h}.
\end{multline*}
\end{theorem}

\begin{lemma}[{Majorizing measure theorem~\cite[Thm.~8.6.1]{vershynin2018hdp}}]
\label{lem:majorizing}
Let $(X_t)_{t\in\mathcal T}$ be a mean-zero Gaussian process on a set $\mathcal T$, and let $\mathsf d(t,s):=(\mathbb E(X_t-X_s)^2)^{1/2}$ be its canonical metric. There are absolute constants $c_-,c_+>0$ such that
\[
c_-\,\gamma_2(\mathcal T,\mathsf d)\le\mathbb E\sup_{t\in\mathcal T}X_t\le c_+\,\gamma_2(\mathcal T,\mathsf d).
\]
In particular, let $D\ge1$, $\mathcal V\subset\mathbb R^D$, and let $\boldsymbol\zeta$ be a standard Gaussian vector in $\mathbb R^D$. The process $X_{\mathbf v}:=\langle\boldsymbol\zeta,\mathbf v\rangle$, $\mathbf v\in\mathcal V$, has canonical metric $\mathsf d(\mathbf u,\mathbf v)=\normtwo{\mathbf u-\mathbf v}$, so, with $c_2:=1/c_-$,
\[
\gamma_2(\mathcal V,\normtwo{\cdot})\le c_2\,\mathbb E\sup_{\mathbf v\in\mathcal V}\langle\boldsymbol\zeta,\mathbf v\rangle.
\]
\end{lemma}

\begin{lemma}
\label{lem:gamma2contract}
Let $(\mathcal T,\mathsf d)$ and $(\mathcal T',\mathsf d')$ be metric spaces, let $\omega\ge0$, and let $\varphi:\mathcal T\to\mathcal T'$ satisfy $\mathsf d'(\varphi(\mathbf x),\varphi(\mathbf y))\le\omega\,\mathsf d(\mathbf x,\mathbf y)$ for all $\mathbf x,\mathbf y\in\mathcal T$. Then
\[
\gamma_2\big(\varphi(\mathcal T),\mathsf d'\big)\le\omega\,\gamma_2(\mathcal T,\mathsf d).
\]
\end{lemma}
Its proof is in Appendix~\ref{app:gamma2}.
\begin{lemma}[{[\citen{foucartrauhut2013}, Prop.~7.29]}]
\label{lem:gaussmax}
Let $X_1,\dots,X_K$ be mean-zero random variables, not necessarily independent, such that $\mathbb E\exp(\theta X_i)\le\exp(\mu\,\theta^2)$ for all $\theta\in\mathbb R$ and $i=1,\dots,K$. Then
\[
\mathbb E\max_{1\le i\le K}|X_i|\le\sqrt{4\mu\log(2K)}.
\]
A standard Gaussian random variable satisfies this hypothesis with $\mu=1/2$~\cite[Lem~7.6]{foucartrauhut2013}.
\end{lemma}

\emph{Reduction to a Gaussian chaos process.} Let $\mathbf S=\blkdiag(\mathbf S_1,\dots,\mathbf S_L)$ be as in Theorem~\ref{thm:mamb}. Throughout Section~\ref{sec:proof-ma}, $m_\ell\ge m_1(\varepsilon,\lambda,\delta)\,\eta_\ell$ for every $\ell$.

Let $\mathcal Y:=\{\mathbf y\in\mathbb R^{d+1}\mid\|\mathbf y\|_1\le1\}$. For $\mathbf y=(\mathbf x,y_{d+1})\in\mathcal Y$, set $\mathbf t(\mathbf y):=[\mathbf A,\mathbf b]\mathbf y=\mathbf A\mathbf x+y_{d+1}\mathbf b$, so that $T=\{\mathbf t(\mathbf y)\mid\mathbf y\in\mathcal Y\}$. Partition $\mathbf t(\mathbf y)$ conformally with the blocks of $\mathbf S$, as $\mathbf t_\ell(\mathbf y)=[\mathbf A_\ell,\mathbf b_\ell]\mathbf y$. Let $\boldsymbol\xi$ be the vector of all entries of the matrices $\sqrt{m_\ell}\,\mathbf S_\ell$, $\ell=1,\dots,L$, which are i.i.d.\ standard Gaussian. Then $\mathbf S\mathbf t(\mathbf y)=\mathbf M(\mathbf y)\boldsymbol\xi$, where
\[
\mathbf M(\mathbf y):=\blkdiag_{\ell=1}^{L}\Big(\tfrac1{\sqrt{m_\ell}}\mathbf I_{m_\ell}\otimes\mathbf t_\ell(\mathbf y)^\top\Big).
\]
Since $\boldsymbol\xi$ is isotropic, $\mathbb E\normtwo{\mathbf S\mathbf t(\mathbf y)}^2=\normtwo{\mathbf t(\mathbf y)}^2$. For $k\ge0$, let
\begin{align*}
T_k&:=\{\mathbf z\in T\mid\normtwo{\mathbf z}\le2^{-k}\},\\
\mathcal Y_k&:=\{\mathbf y\in\mathcal Y\mid\mathbf t(\mathbf y)\in T_k\},\\
\mathcal A_k&:=\{\mathbf M(\mathbf y)\mid\mathbf y\in\mathcal Y_k\}.
\end{align*}
Then
\begin{equation}
\sup_{\mathbf z\in T_k}\big|\normtwo{\mathbf S\mathbf z}^2-\normtwo{\mathbf z}^2\big|=\sup_{\mathbf M\in\mathcal A_k}\big|\normtwo{\mathbf M\boldsymbol\xi}^2-\mathbb E\normtwo{\mathbf M\boldsymbol\xi}^2\big|, \label{eq:chaos}
\end{equation}
the supremum of the centered Gaussian chaos $\boldsymbol\xi^\top\mathbf M^\top\mathbf M\boldsymbol\xi-\operatorname{tr}(\mathbf M^\top\mathbf M)$ over $\mathbf M\in\mathcal A_k$~\cite{krahmermendelsonrauhut2014,dirksen2015}, to which Theorem~\ref{thm:dirksen} applies. The entries of $\boldsymbol\xi$ are standard Gaussian, so $\|\boldsymbol\xi\|_{\psi_2}$ is an absolute constant, which we absorb into $c_7$ and $c_8$. Every point of $T$ has norm at most $1$, so $T=T_0\supset T_1\supset T_2\supset\cdots$. For $k\ge0$, let $R_k:=T_k\setminus T_{k+1}$, so that every $\mathbf z\in R_k$ satisfies $2^{-(k+1)}<\normtwo{\mathbf z}\le2^{-k}$. As in~\cite[Sec.~3.1]{narayanannelson2019}, we bound the deviation on each $T_k$ separately, use this bound on $R_k\subset T_k$, and then take a union bound over $k$.

\emph{Complexity parameters.} The next lemma bounds the quantities in Theorem~\ref{thm:dirksen} for $\mathcal A_k$ in terms of $m_1(\varepsilon,\lambda,\delta)$.

\begin{lemma}
\label{lem:complexity}
If $m_\ell\ge m_1(\varepsilon,\lambda,\delta)\,\eta_\ell$ for every $\ell$, then the following hold for every $k\ge0$.

\titem{i}{lem:frob}$\bar\Delta_2(\mathcal A_k)\le2^{-k}$.
\titem{ii}{lem:spec}$\bar\Delta_\infty(\mathcal A_k)\le2/\sqrt{m_1(\varepsilon,\lambda,\delta)}$.
\titem{iii}{lem:schatten4}$\bar\Delta_4(\mathcal A_k)^2\le2\cdot2^{-k}/\sqrt{m_1(\varepsilon,\lambda,\delta)}$.
\titem{iv}{lem:entropy}$\gamma_2(\mathcal A_k,\mathsf d_\infty)\le\dfrac{2c_2}{\sqrt{m_1(\varepsilon,\lambda,\delta)}}\sqrt{2\log(2(d+1))}$.
\end{lemma}

\begin{proof}
Fix $\mathbf y\in\mathcal Y$. The $\ell$-th diagonal block $\tfrac1{\sqrt{m_\ell}}\mathbf I_{m_\ell}\otimes\mathbf t_\ell(\mathbf y)^\top$ of $\mathbf M(\mathbf y)$ has $m_\ell$ singular values equal to $\normtwo{\mathbf t_\ell(\mathbf y)}/\sqrt{m_\ell}$, and all its other singular values are zero. Since $\normtwo{\mathbf y}\le\|\mathbf y\|_1\le1$, we also have $\normtwo{\mathbf t_\ell(\mathbf y)}\le\normtwo{[\mathbf A_\ell,\mathbf b_\ell]}=2\sqrt{\eta_\ell}$. Since $m_\ell\ge m_1(\varepsilon,\lambda,\delta)\,\eta_\ell$,
\begin{equation}
\frac{\normtwo{[\mathbf A_\ell,\mathbf b_\ell]}}{\sqrt{m_\ell}}=\frac{2\sqrt{\eta_\ell}}{\sqrt{m_\ell}}\le\frac2{\sqrt{m_1(\varepsilon,\lambda,\delta)}},\qquad\ell=1,\dots,L. \label{eq:blockratio}
\end{equation}

\emph{(i)} $\normF{\mathbf M(\mathbf y)}=\normtwo{\mathbf t(\mathbf y)}$, which is at most $2^{-k}$ when $\mathbf y\in\mathcal Y_k$.

\emph{(ii)} $\normtwo{\mathbf M(\mathbf y)}=\max_\ell\normtwo{\mathbf t_\ell(\mathbf y)}/\sqrt{m_\ell}\le\max_\ell\normtwo{[\mathbf A_\ell,\mathbf b_\ell]}/\sqrt{m_\ell}\le2/\sqrt{m_1(\varepsilon,\lambda,\delta)}$, by~\eqref{eq:blockratio}.

\emph{(iii)} Let $\mathbf y\in\mathcal Y_k$. Then
\begin{tcmultline*}
\normSp{\mathbf M(\mathbf y)}{4}^4\eqstep{a}\sum_{\ell=1}^L\frac{\normtwo{\mathbf t_\ell(\mathbf y)}^4}{m_\ell}\lestep{b}\frac1{m_1(\varepsilon,\lambda,\delta)}\sum_{\ell=1}^L\frac{\normtwo{\mathbf t_\ell(\mathbf y)}^4}{\eta_\ell} \tcbr
\lestep{c}\frac4{m_1(\varepsilon,\lambda,\delta)}\sum_{\ell=1}^L\normtwo{\mathbf t_\ell(\mathbf y)}^2\lestep{d}\frac{4\cdot2^{-2k}}{m_1(\varepsilon,\lambda,\delta)}.
\end{tcmultline*}
Step~(a) uses the singular values of the blocks. Step~(b) uses $m_\ell\ge m_1(\varepsilon,\lambda,\delta)\,\eta_\ell$. Step~(c) uses $\normtwo{\mathbf t_\ell(\mathbf y)}^2\le4\eta_\ell$. Step~(d) uses $\sum_\ell\normtwo{\mathbf t_\ell(\mathbf y)}^2=\normtwo{\mathbf t(\mathbf y)}^2\le2^{-2k}$. Taking square roots and the supremum over $\mathcal Y_k$ gives~(iii).

\emph{(iv)} For $\mathbf y,\mathbf y'\in\mathcal Y_k$, the matrix $\mathbf M(\mathbf y)-\mathbf M(\mathbf y')$ is block-diagonal, and, by~\eqref{eq:blockratio}, its $\ell$-th block has spectral norm $\normtwo{[\mathbf A_\ell,\mathbf b_\ell](\mathbf y-\mathbf y')}/\sqrt{m_\ell}\le(2/\sqrt{m_1(\varepsilon,\lambda,\delta)})\normtwo{\mathbf y-\mathbf y'}$. Thus $\mathbf y\mapsto\mathbf M(\mathbf y)$ is $(2/\sqrt{m_1(\varepsilon,\lambda,\delta)})$-Lipschitz from $(\mathcal Y_k,\normtwo{\cdot})$ to $(\mathcal A_k,\mathsf d_\infty)$, and Lemma~\ref{lem:gamma2contract} gives $\gamma_2(\mathcal A_k,\mathsf d_\infty)\le(2/\sqrt{m_1(\varepsilon,\lambda,\delta)})\,\gamma_2(\mathcal Y_k,\normtwo{\cdot})$. Let $\boldsymbol\zeta\sim\mathcal N(0,\mathbf I_{d+1})$. By Lemma~\ref{lem:majorizing}, $\mathcal Y_k\subset\mathcal Y$, the duality of the $\ell_1$ and $\ell_\infty$ norms, and Lemma~\ref{lem:gaussmax} with $K=d+1$ and $\mu=1/2$,
\[
\gamma_2(\mathcal Y_k,\normtwo{\cdot})\le c_2\,\mathbb E\sup_{\mathbf y\in\mathcal Y}\langle\boldsymbol\zeta,\mathbf y\rangle=c_2\,\mathbb E\max_{1\le i\le d+1}|\zeta_i|\le c_2\sqrt{2\log(2(d+1))}.
\]
Combining the two bounds gives~(iv).
\end{proof}

Let
\begin{gather*}
c_9:=\Big(2+\frac1{2c_2^2}\Big)^{-1},\qquad c_{10}:=c_7+\frac{c_8}{c_9},\\
c_{11}:=\max\big\{6c_{10},\ 1+\sqrt{1+2c_{10}}\big\},
\end{gather*}
let $\varepsilon':=\varepsilon/c_{11}$, and let
\begin{multline}
h:=\log\Big(\frac{(2d+2)\log(2/(\varepsilon'\lambda))}{\delta}\Big) \\
=\log(1/\delta)+\log(2d+2)+\log\log(2/(\varepsilon'\lambda)), \label{eq:hdef}
\end{multline}
which satisfies $h\ge\log4+\log\log2>1$. The next lemma applies Theorem~\ref{thm:dirksen} on each $T_k$.

\begin{lemma}
\label{lem:shell}
If $m_\ell\ge m_1(\varepsilon,\lambda,\delta)\,\eta_\ell$ for every $\ell$, then, for every $k\ge0$, with probability at least $1-e^{-h}$,
\begin{equation}
\sup_{\mathbf z\in T_k}\big|\normtwo{\mathbf S\mathbf z}^2-\normtwo{\mathbf z}^2\big|\le c_{10}\big((\varepsilon'\lambda)^2+\varepsilon'\lambda\cdot2^{-k}\big). \label{eq:requirement}
\end{equation}
\end{lemma}

Its proof applies Theorem~\ref{thm:dirksen} to $\mathcal A_k$, with~\eqref{eq:chaos} and the bounds of Lemma~\ref{lem:complexity}, and is in Appendix~\ref{app:shell}.

\begin{proof}[Proof of Theorem~\ref{thm:mamb}(i)]
This proof is similar to that in~\cite[proof of Thm.~3.1]{narayanannelson2019}.
Let $k_0:=\lceil\log_2(1/(\varepsilon'\lambda))\rceil$. Since $T=T_0\supset T_1\supset\cdots$, the set $T$ is the disjoint union
\[
T=\bigcup_{k=0}^{k_0-1}R_k\ \cup\ T_{k_0}.
\]
By Lemma~\ref{lem:shell}, for each $0\le k\le k_0$, the bound~\eqref{eq:requirement} fails with probability at most $e^{-h}$. Since $k_0+1\le(2d+2)\log(2/(\varepsilon'\lambda))$, a union bound over $0\le k\le k_0$ shows that~\eqref{eq:requirement} holds for every $0\le k\le k_0$ with probability at least $1-(k_0+1)e^{-h}\ge1-\delta$.

On this event, let $\mathbf z\in T$. If $\mathbf z\in R_k$ for some $k<k_0$, then $R_k\subset T_k$, $\normtwo{\mathbf z}>2^{-(k+1)}$ and $2^k\varepsilon'\lambda<1$, so
\begin{tcmultline*}
\big|\normtwo{\mathbf S\mathbf z}-\normtwo{\mathbf z}\big|=\frac{|\normtwo{\mathbf S\mathbf z}^2-\normtwo{\mathbf z}^2|}{\normtwo{\mathbf S\mathbf z}+\normtwo{\mathbf z}} \tcbr
\le2^{k+1}c_{10}\big((\varepsilon'\lambda)^2+\varepsilon'\lambda\cdot2^{-k}\big)\le6c_{10}\,\varepsilon'\lambda.
\end{tcmultline*}
If $\mathbf z\in T_{k_0}$, then $\normtwo{\mathbf z}\le2^{-k_0}\le\varepsilon'\lambda$, so~\eqref{eq:requirement} gives $\normtwo{\mathbf S\mathbf z}^2\le\normtwo{\mathbf z}^2+2c_{10}(\varepsilon'\lambda)^2\le(1+2c_{10})(\varepsilon'\lambda)^2$. The triangle inequality then gives $\big|\normtwo{\mathbf S\mathbf z}-\normtwo{\mathbf z}\big|\le\big(1+\sqrt{1+2c_{10}}\big)\varepsilon'\lambda$. In both cases, $\big|\normtwo{\mathbf S\mathbf z}-\normtwo{\mathbf z}\big|\le c_{11}\,\varepsilon'\lambda=\varepsilon\lambda$. Hence $\mathbf S$ satisfies C1$(\varepsilon,\lambda)$ with probability at least $1-\delta$.
\end{proof}

\subsubsection{Proof of Theorem~\ref{thm:mamb}(ii)}
\label{sec:proof-mb}

\begin{proof}
If $\mathbf b=\mathbf 0$, C2$(\lambda)$ holds trivially. Otherwise, let $\mathbf U:=\mathbf b/\normtwo{\mathbf b}$ (an $n\times1$ matrix with orthonormal columns), so that $\Gamma(\mathbf U_\ell)\le\normtwo{\mathbf U_\ell}^2=\beta_\ell$. Let $c_4:=c_1^2$. Then $m_2(\lambda,\delta)=c_1^2\log(2/\delta)/(\lambda/3)^2$ by~\eqref{eq:mbfun}, and $m_2(\lambda,\delta)\,\Gamma(\mathbf U_\ell)\le m_2(\lambda,\delta)\,\beta_\ell\le m_\ell$ for every $\ell$. Lemma~\ref{lem:srinivasa} with $r=1$, $\varepsilon'=\lambda/3$ and $M_0=m_2(\lambda,\delta)$ therefore gives $|\normtwo{\mathbf S\mathbf b}^2-\normtwo{\mathbf b}^2|\le\tfrac{\lambda}{3}\normtwo{\mathbf b}^2$ with probability at least $1-\delta$.
\end{proof}


\subsection{Proof of Theorem~\ref{thm:main}}
\label{sec:mainproof}

\begin{proof}
By~\eqref{eq:allocation-main} and Theorem~\ref{thm:mamb}, C1$(\varepsilon,\lambda)$ and C2$(\lambda)$ each fail with probability at most $\delta/2$. By a union bound, both hold with probability at least $1-\delta$. On this event, Theorem~\ref{thm:partI} gives $F(\widetilde{\mathbf x})\le(1+O(\varepsilon))\,\OPT$. For $m_\ell=\big\lceil\max\{m_1(\varepsilon,\lambda,\delta/2)\,\eta_\ell,\ m_2(\lambda,\delta/2)\,\beta_\ell\}\big\rceil$, bounding the maximum by the sum and summing over $\ell$ gives $m=O\big(m_1(\varepsilon,\lambda,\delta/2)\,\eta_\Sigma+m_2(\lambda,\delta/2)\big)$, since $\sum_\ell\beta_\ell\le1$.
\end{proof}
\subsection{Proof of Theorem~\ref{thm:transfer}}
\label{sec:twostage}
The proof of part~(i) has two steps. By C1$(\varepsilon/2,\lambda)$, the first stage $\widehat{\mathbf S}$ distorts the norm of each point of $T$ by at most $\varepsilon\lambda/2$. By Theorem~\ref{thm:nn}, the second stage $\widetilde{\mathbf S}$ distorts the norm of each point of $\widehat{\mathbf S}(T)$ by at most $\varepsilon\lambda/2$. The triangle inequality then gives C1$(\varepsilon,\lambda)$ for $\widetilde{\mathbf S}\widehat{\mathbf S}$. In part~(ii), C2$(3\lambda/7)$ for $\widehat{\mathbf S}$ and Theorem~\ref{thm:woodruffjl} with parameter $\lambda/7$ for $\widetilde{\mathbf S}$ together give a factor between $(1-\lambda/7)^2$ and $(1+\lambda/7)^2$, and both lie between $1-\lambda/3$ and $1+\lambda/3$. The proof uses the following three results.

In~\cite[Sec.~3.1]{narayanannelson2019}, $\mathcal X=\{\mathbf x_1,\dots,\mathbf x_N\}\subset\mathbb R^D$ is a set of unit vectors, $\mathcal C:=\mathrm{conv}(\mathcal X)$, $\boldsymbol\Pi=\boldsymbol\Pi_0/\sqrt p$, where $\boldsymbol\Pi_0\in\mathbb R^{p\times D}$ has i.i.d.\ sub-Gaussian entries with mean~0 and variance~1, and $\mathcal C_j:=\{\mathbf c\in\mathcal C\mid\normtwo{\mathbf c}\le2^{-j}\}$.

\begin{theorem}[{[\citen{narayanannelson2019}, Thm.~3.1]}]
\label{thm:nn}
Let $\varepsilon',\delta'\in(0,1)$ and $p=\Theta\big(\varepsilon'^{-2}\log(N\log(2/\varepsilon')/\delta')\big)$. There is an absolute constant $\bar c$ such that, for all $j\ge0$,
\begin{tcmultline*}
\mathbb P\Big(\sup_{\mathbf c\in\mathcal C_j}\big|\normtwo{\boldsymbol\Pi\mathbf c}^2-\normtwo{\mathbf c}^2\big|\ge \bar c\big(\varepsilon'^2+\varepsilon'\cdot2^{-j}\big)\Big) \tcbr
\le\frac{\delta'}{N\log(2/\varepsilon')}.
\end{tcmultline*}
Consequently, with probability at least $1-\delta'/N$, $\big|\normtwo{\boldsymbol\Pi\mathbf c}-\normtwo{\mathbf c}\big|=O(\varepsilon')$ for all $\mathbf c\in\mathcal C$.
\end{theorem}

We write $c_{12}$ for the absolute constant in the conclusion $O(\varepsilon')$ of Theorem~\ref{thm:nn}.

\begin{definition}[{[\citen{woodruff2014}, Def.~3]}]
\label{def:jlt}
A random matrix $\boldsymbol\Phi\in\mathbb R^{p\times D}$ forms a Johnson--Lindenstrauss transform with parameters $\varepsilon',\delta',f$, or JLT$(\varepsilon',\delta',f)$ for short, if with probability at least $1-\delta'$, for any $f$-element subset $\mathcal W\subset\mathbb R^D$, for all $\mathbf w,\mathbf w'\in\mathcal W$, $|\langle\boldsymbol\Phi\mathbf w,\boldsymbol\Phi\mathbf w'\rangle-\langle\mathbf w,\mathbf w'\rangle|\le\varepsilon'\normtwo{\mathbf w}\normtwo{\mathbf w'}$.
\end{definition}

\begin{theorem}[{[\citen{woodruff2014}, Thm.~4]}]
\label{thm:woodruffjl}
Let $\varepsilon',\delta'\in(0,1)$, and let $\boldsymbol\Phi=\mathbf R/\sqrt p\in\mathbb R^{p\times D}$, where the entries of $\mathbf R$ are independent standard normal random variables. If $p=\Omega(\varepsilon'^{-2}\log(f/\delta'))$, then $\boldsymbol\Phi$ is a JLT$(\varepsilon',\delta',f)$.
\end{theorem}

\begin{proof}[Proof of Theorem~\ref{thm:transfer}(i)]
\emph{Step 1.} Let $\mathbf h_i:=\widehat{\mathbf S}\mathbf a_i$ for $i\le d$, and $\mathbf h_{d+1}:=\widehat{\mathbf S}\mathbf b$, and let $V$ be the set of the vectors $\pm\mathbf h_j/\normtwo{\mathbf h_j}$ over $\mathbf h_j\ne\mathbf 0$, so that $|V|\le2d+2$. We show that $\widehat{\mathbf S}(T)\subset\tfrac32\,\mathrm{conv}(V)$. Since $\mathbf a_i,\mathbf b\in T$ and $\normtwo{\mathbf a_i},\normtwo{\mathbf b}\le1$, C1$(\varepsilon/2,\lambda)$ gives $\normtwo{\mathbf h_j}\le1+\varepsilon\lambda/2\le3/2$. Then $\mathbf 0\in\tfrac32\mathrm{conv}(V)$, as the midpoint of $\pm\tfrac32\mathbf v$ for any $\mathbf v\in V$. Each $\pm\mathbf h_j\ne\mathbf 0$ is the convex combination $\pm\mathbf h_j=\tfrac{2\normtwo{\mathbf h_j}}{3}\cdot\tfrac32\tfrac{\pm\mathbf h_j}{\normtwo{\mathbf h_j}}+\big(1-\tfrac{2\normtwo{\mathbf h_j}}{3}\big)\cdot\mathbf 0$ of two points of the convex set $\tfrac32\mathrm{conv}(V)$. Since $\widehat{\mathbf S}(T)=\mathrm{conv}\{\pm\mathbf h_1,\dots,\pm\mathbf h_{d+1}\}$, we obtain $\widehat{\mathbf S}(T)\subset\tfrac32\,\mathrm{conv}(V)$.

\emph{Step 2.} Conditionally on $\widehat{\mathbf S}$, the set $V$ is fixed, and $\widetilde{\mathbf S}$ is independent of it. We apply Theorem~\ref{thm:nn} with
\begin{gather*}
\mathcal X=V,\quad N=|V|\le2d+2,\quad D=m,\quad\boldsymbol\Pi=\widetilde{\mathbf S},\\
p=\widetilde m,\quad\delta'=\delta,\quad\varepsilon'=\varepsilon\lambda/(3c_{12}).
\end{gather*}
Its condition on the sketch size is $\widetilde m\ge\widetilde m_1(\varepsilon,\lambda,\delta)$, with $c_5$ absorbing $c_{12}$. With probability at least $1-\delta/|V|\ge1-\delta$, every $\mathbf v\in\mathrm{conv}(V)$ satisfies $|\normtwo{\widetilde{\mathbf S}\mathbf v}-\normtwo{\mathbf v}|\le\varepsilon\lambda/3$. Let $\mathbf w\in\widehat{\mathbf S}(T)$. By Step~1, $\mathbf w=\tfrac32\mathbf v$ for some $\mathbf v\in\mathrm{conv}(V)$. Hence, for all $\mathbf w\in\widehat{\mathbf S}(T)$,
\[
\big|\normtwo{\widetilde{\mathbf S}\mathbf w}-\normtwo{\mathbf w}\big|=\tfrac32\big|\normtwo{\widetilde{\mathbf S}\mathbf v}-\normtwo{\mathbf v}\big|\le\frac{\varepsilon\lambda}{2}.
\]

\emph{Step 3.} For every $\mathbf z\in T$,
\begin{tcmultline*}
\big|\normtwo{\widetilde{\mathbf S}\widehat{\mathbf S}\mathbf z}-\normtwo{\mathbf z}\big|\lestep{a}\big|\normtwo{\widetilde{\mathbf S}\widehat{\mathbf S}\mathbf z}-\normtwo{\widehat{\mathbf S}\mathbf z}\big|+\big|\normtwo{\widehat{\mathbf S}\mathbf z}-\normtwo{\mathbf z}\big| \tcbr
\lestep{b}\frac{\varepsilon\lambda}{2}+\frac{\varepsilon\lambda}{2}=\varepsilon\lambda.
\end{tcmultline*}
Step~(a) is the triangle inequality. Step~(b) uses Step~2 with $\mathbf w=\widehat{\mathbf S}\mathbf z\in\widehat{\mathbf S}(T)$, and C1$(\varepsilon/2,\lambda)$. This is C1$(\varepsilon,\lambda)$ for $\widetilde{\mathbf S}\widehat{\mathbf S}$.
\end{proof}

\begin{proof}[Proof of Theorem~\ref{thm:transfer}(ii)]
If $\mathbf b=\mathbf 0$, C2$(\lambda)$ holds trivially. Let $\mathbf b\ne\mathbf 0$. Conditionally on $\widehat{\mathbf S}$, the vector $\widehat{\mathbf S}\mathbf b$ is fixed, and $\widetilde{\mathbf S}$ is independent of it. We apply Theorem~\ref{thm:woodruffjl} with
\[
\boldsymbol\Phi=\widetilde{\mathbf S},\quad p=\widetilde m,\quad D=m,\quad f=1,\quad\varepsilon'=\lambda/7,\quad\delta'=\delta.
\]
Its condition on the sketch size is $\widetilde m\ge\widetilde m_2(\lambda,\delta)$. Taking $\mathcal W=\{\widehat{\mathbf S}\mathbf b\}$ and $\mathbf w=\mathbf w'=\widehat{\mathbf S}\mathbf b$ in Definition~\ref{def:jlt} gives, with probability at least $1-\delta$,
\[
\big|\normtwo{\widetilde{\mathbf S}\widehat{\mathbf S}\mathbf b}^2-\normtwo{\widehat{\mathbf S}\mathbf b}^2\big|\le\frac{\lambda}{7}\normtwo{\widehat{\mathbf S}\mathbf b}^2.
\]
By C2$(3\lambda/7)$, $(1-\frac\lambda7)\normtwo{\mathbf b}^2\le\normtwo{\widehat{\mathbf S}\mathbf b}^2\le(1+\frac\lambda7)\normtwo{\mathbf b}^2$. Combining the two bounds,
\begin{tcmultline*}
\Big(1-\frac{\lambda}{3}\Big)\normtwo{\mathbf b}^2\lestep{a}\Big(1-\frac{\lambda}{7}\Big)^2\normtwo{\mathbf b}^2\le\normtwo{\widetilde{\mathbf S}\widehat{\mathbf S}\mathbf b}^2 \tcbr
\le\Big(1+\frac{\lambda}{7}\Big)^2\normtwo{\mathbf b}^2\lestep{b}\Big(1+\frac{\lambda}{3}\Big)\normtwo{\mathbf b}^2.
\end{tcmultline*}
Step~(a) uses $(1-\lambda/7)^2\ge1-2\lambda/7$. Step~(b) uses $(1+\lambda/7)^2=1+2\lambda/7+\lambda^2/49\le1+15\lambda/49$, since $\lambda<1$. This is C2$(\lambda)$ for $\widetilde{\mathbf S}\widehat{\mathbf S}$.
\end{proof}


\subsection{Proof of Theorem~\ref{thm:twostage}}
\label{sec:twostageproof}

\begin{proof}[Proof of Theorem~\ref{thm:twostage}]
By Theorem~\ref{thm:mamb}(i) at $(\varepsilon/2,\lambda,\delta/4)$ and Theorem~\ref{thm:mamb}(ii) at $(3\lambda/7,\delta/4)$, the matrix $\widehat{\mathbf S}$ satisfies C1$(\varepsilon/2,\lambda)$ and C2$(3\lambda/7)$, each with probability at least $1-\delta/4$. Let $E$ be the event that both hold. Then $\mathbb P(E^c)\le\delta/2$. On $E$, Theorem~\ref{thm:transfer} at failure probability $\delta/4$ and a union bound show that, conditionally on $\widehat{\mathbf S}$, the matrix $\widetilde{\mathbf S}\widehat{\mathbf S}$ satisfies C1$(\varepsilon,\lambda)$ and C2$(\lambda)$ with probability at least $1-\delta/2$. Hence $\mathbf S=\widetilde{\mathbf S}\widehat{\mathbf S}$ satisfies C1$(\varepsilon,\lambda)$ and C2$(\lambda)$ with probability at least $1-\delta$. On this event, Theorem~\ref{thm:partI} gives $F(\widetilde{\mathbf x})\le(1+1906\,\varepsilon)\,\OPT$.

The order of $\widetilde m$ follows from~\eqref{eq:mtildea} and~\eqref{eq:mtildeb}, since $\widetilde m_2(\lambda,\delta/4)=O\big(\widetilde m_1(\varepsilon,\lambda,\delta/4)\big)$.
\end{proof}

\subsection{Estimating the Local Spectral Norms}
\label{sec:etaest}

This subsection shows that each node can estimate $\eta_\ell$ from its own data. Theorems~\ref{thm:mamb} and~\ref{thm:twostage} use $\eta_\ell$ in the sketch size of node $\ell$, and an estimate within a factor of~$2$ suffices.

We use block Krylov iteration. The work~\cite{muscomusco2015} gives two algorithms, Simultaneous Iteration~\cite[Alg.~1]{muscomusco2015} and Block Krylov Iteration~\cite[Alg.~2]{muscomusco2015}. Each takes a matrix $\mathbf B\in\mathbb R^{p\times q}$, an error parameter $\tau$ and a rank $\kappa$, and returns a matrix $\mathbf Z$ with $\kappa$ orthonormal columns $\mathbf z_1,\dots,\mathbf z_\kappa$. Let $\sigma_1\ge\sigma_2\ge\cdots$ be the singular values of $\mathbf B$, and let $\mathbf w_1,\mathbf w_2,\dots$ be its left singular vectors.

\begin{theorem}[{Per Vector Quality Guarantee~\cite[Thm.~12]{muscomusco2015}}]
\label{thm:mm12}
With probability $99/100$, \cite[Alg.~1]{muscomusco2015} and~\cite[Alg.~2]{muscomusco2015} return $\mathbf Z$ satisfying
\[
\big|\mathbf w_i^\top\mathbf B\mathbf B^\top\mathbf w_i-\mathbf z_i^\top\mathbf B\mathbf B^\top\mathbf z_i\big|\le\tau\sigma_{\kappa+1}^2\quad\text{for all }i.
\]
\end{theorem}

\begin{theorem}[{Block Krylov Iteration Runtime~\cite[Thm.~7]{muscomusco2015}}]
\label{thm:mm7}
\cite[Alg.~2]{muscomusco2015} runs in time
\begin{tcmultline*}
O\Big(\nnz(\mathbf B)\frac{\kappa\log q}{\sqrt\tau}+\frac{p\kappa^2\log^2q}{\tau} \tcbr
+\frac{\kappa^3\log^3q}{\tau^{3/2}}\Big).
\end{tcmultline*}
\end{theorem}

\begin{lemma}
\label{lem:etaest}
Fix a node $\ell$ and $\delta\in(0,1)$. Run~\cite[Alg.~2]{muscomusco2015} on $[\mathbf A_\ell,\mathbf b_\ell]\in\mathbb R^{n_\ell\times(d+1)}$ with $\kappa=1$ and $\tau=1/2$, independently $\Nrun:=\lceil\log_{100}(2L/\delta)\rceil$ times. Let $\mathbf z^{(1)},\dots,\mathbf z^{(\Nrun)}$ be the returned unit vectors, and let
\[
\widehat\eta_\ell:=\tfrac12\max_{1\le s\le\Nrun}\normtwo{[\mathbf A_\ell,\mathbf b_\ell]^\top\mathbf z^{(s)}}^2.
\]
Then, with probability at least $1-\delta/(2L)$, $\eta_\ell\le\widehat\eta_\ell\le2\eta_\ell$. For dense $[\mathbf A_\ell,\mathbf b_\ell]$, the cost is $O\big(n_\ell\,d\log(d+1)\log(L/\delta)\big)$ flops.
\end{lemma}
\begin{proof}
Let $\sigma_1\ge\sigma_2$ be the two largest singular values of $[\mathbf A_\ell,\mathbf b_\ell]$, and let $\mathbf w_1$ be a leading left singular vector, so that $\sigma_1^2=4\eta_\ell$. For every unit vector $\mathbf z$, $\normtwo{[\mathbf A_\ell,\mathbf b_\ell]^\top\mathbf z}^2\le\sigma_1^2=4\eta_\ell$. Hence $\widehat\eta_\ell\le2\eta_\ell$ always. For a single run, Theorem~\ref{thm:mm12} with $\mathbf B=[\mathbf A_\ell,\mathbf b_\ell]$, $\kappa=1$, $i=1$ and $\tau=1/2$ gives, with probability $99/100$,
\begin{tcmultline*}
\normtwo{[\mathbf A_\ell,\mathbf b_\ell]^\top\mathbf z^{(s)}}^2\ge\mathbf w_1^\top[\mathbf A_\ell,\mathbf b_\ell][\mathbf A_\ell,\mathbf b_\ell]^\top\mathbf w_1-\tfrac12\sigma_2^2 \tcbr
=\sigma_1^2-\tfrac12\sigma_2^2\ge\tfrac12\sigma_1^2=2\eta_\ell.
\end{tcmultline*}
If this holds for at least one run, then $\widehat\eta_\ell\ge\eta_\ell$. The runs are independent, so all $\Nrun$ runs fail with probability at most $100^{-\Nrun}\le\delta/(2L)$.

For the cost, Theorem~\ref{thm:mm7} with $p=n_\ell$, $q=d+1$, $\kappa=1$, $\tau=1/2$ and $\nnz([\mathbf A_\ell,\mathbf b_\ell])=n_\ell(d+1)$ gives $O(n_\ell\,d\log(d+1))$ per run. Computing $\normtwo{[\mathbf A_\ell,\mathbf b_\ell]^\top\mathbf z^{(s)}}^2$ adds $O(n_\ell d)$. Multiplying by $\Nrun=O(\log(L/\delta))$ gives the claim.
\end{proof}

A union bound over the $L$ nodes makes all the estimates valid with probability at least $1-\delta/2$. Theorems~\ref{thm:main} and~\ref{thm:twostage} use $\eta_\ell$ only through $m_\ell\ge m_1(\varepsilon,\lambda,\delta/2)\,\eta_\ell$ and $m_\ell\ge m_1(\varepsilon/2,\lambda,\delta/4)\,\eta_\ell$. On this event, the choice $m_\ell=\big\lceil\max\{m_1(\varepsilon/2,\lambda,\delta/8)\,\widehat\eta_\ell,\ m_2(3\lambda/7,\delta/8)\,\beta_\ell\}\big\rceil$ therefore satisfies the hypothesis of Theorem~\ref{thm:twostage} with $\delta/2$ in place of $\delta$. Its first term is at most twice its value with the exact $\eta_\ell$. By a union bound, the estimates and the guarantee of Theorem~\ref{thm:twostage} hold together with probability at least $1-\delta$.

\section{Conclusion}
\label{sec:conclusion}

This paper proposed communication-efficient sketching for sparse linear regression when the data is distributed among nodes that communicate with a central node. We showed that a block-diagonal Gaussian sketch, in which each node compresses only its own rows with a sketch size determined by its local data, attains the relative-error guarantee of the centralized sketch. Its total sketch size scales with a heterogeneity measure of the data across the nodes, rather than with the number of nodes. A Gaussian sketch applied at the central node to the stack of locally sketched data reduces the sketch size to the order of that of the centralized sketch, without further communication, and attains the same guarantee. Up to lower-order terms, the communication cost of this two-stage sketch scales with the heterogeneity measure, while that of the distributed implementation of the centralized sketch scales with the number of nodes. For Gaussian data split equally across the nodes, the heterogeneity measure is bounded by a constant for all sufficiently large numbers of nodes, almost surely, so the communication cost does not grow with the number of nodes, up to lower-order terms. When the required accuracy is high, each node holds many rows, and the heterogeneity measure is much smaller than the number of nodes, the two-stage sketch also requires less computation than the distributed implementation of the centralized sketch.

\appendices


\section{Proof of Lemma~\ref{lem:shell}}
\label{app:shell}

Throughout, $m_\ell\ge m_1(\varepsilon,\lambda,\delta)\,\eta_\ell$ for every $\ell$, and $c_9$, $c_{10}$, $c_{11}$, $\varepsilon'$ and $h$ are as in Section~\ref{sec:proof-ma}. Let $a:=\log(1+\log_2c_{11})$. The absolute constant in~\eqref{eq:mafun} is
\[
c_3:=4c_{11}^2\Big(1+\frac{a}{2c_2^2\log4}\Big).
\]
We first bound $h$ and $\log(2(d+1))$ in terms of $m_1(\varepsilon,\lambda,\delta)$. By~\eqref{eq:mafun}, each of $2c_2^2\log(2(d+1))$, $\log(1/\delta)$ and $\log\log(2/(\varepsilon\lambda))$ is at most $\varepsilon^2\lambda^2m_1(\varepsilon,\lambda,\delta)/c_3$, and, since $d\ge1$, $\varepsilon^2\lambda^2m_1(\varepsilon,\lambda,\delta)/c_3\ge2c_2^2\log4$. Since $\log(2/(\varepsilon\lambda))\ge\log2$, we have $\log(2c_{11}/(\varepsilon\lambda))\le(1+\log_2c_{11})\log(2/(\varepsilon\lambda))$, and hence $\log\log(2/(\varepsilon'\lambda))\le\log\log(2/(\varepsilon\lambda))+a$. The three terms of~\eqref{eq:hdef} therefore satisfy
\begin{align*}
\log(1/\delta)&\le\frac{\varepsilon^2\lambda^2m_1(\varepsilon,\lambda,\delta)}{c_3},\qquad\log(2d+2)\le\frac{\varepsilon^2\lambda^2m_1(\varepsilon,\lambda,\delta)}{2c_2^2c_3},\\
\log\log\frac{2}{\varepsilon'\lambda}&\le\Big(1+\frac{a}{2c_2^2\log4}\Big)\frac{\varepsilon^2\lambda^2m_1(\varepsilon,\lambda,\delta)}{c_3}.
\end{align*}
Summing, and using the definitions of $c_9$, $c_3$ and $\varepsilon'=\varepsilon/c_{11}$,
\begin{tcmultline}
h\le\Big(2+\frac1{2c_2^2}\Big)\Big(1+\frac{a}{2c_2^2\log4}\Big)\frac{\varepsilon^2\lambda^2m_1(\varepsilon,\lambda,\delta)}{c_3} \tcbr
=\frac{(\varepsilon'\lambda)^2\,m_1(\varepsilon,\lambda,\delta)}{4c_9}. \label{eq:hm1}
\end{tcmultline}
In the same way,
\begin{equation}
2c_2^2\log(2(d+1))\le\frac{\varepsilon^2\lambda^2m_1(\varepsilon,\lambda,\delta)}{c_3}\le\frac{(\varepsilon'\lambda)^2\,m_1(\varepsilon,\lambda,\delta)}{4}. \label{eq:dm1}
\end{equation}

\begin{proof}[Proof of Lemma~\ref{lem:shell}]
Fix $k\ge0$, and let $\bar\gamma:=\gamma_2(\mathcal A_k,\mathsf d_\infty)$. By Lemma~\ref{lem:complexity} and~\eqref{eq:dm1},
\begin{align*}
\bar\gamma&\le\frac{2}{\sqrt{m_1(\varepsilon,\lambda,\delta)}}\sqrt{2c_2^2\log(2(d+1))}\le\varepsilon'\lambda,\qquad\bar\Delta_2(\mathcal A_k)\le2^{-k},\\
\bar\Delta_\infty(\mathcal A_k)&\le\frac2{\sqrt{m_1(\varepsilon,\lambda,\delta)}},\qquad\bar\Delta_4(\mathcal A_k)^2\le\frac{2\cdot2^{-k}}{\sqrt{m_1(\varepsilon,\lambda,\delta)}}.
\end{align*}
By Theorem~\ref{thm:dirksen} with $h$ as in~\eqref{eq:hdef} and~\eqref{eq:chaos}, with probability at least $1-e^{-h}$,
\begin{tcmultline*}
\sup_{\mathbf z\in T_k}\big|\normtwo{\mathbf S\mathbf z}^2-\normtwo{\mathbf z}^2\big|\le c_7\big(\bar\gamma^2+2^{-k}\bar\gamma\big) \tcbr
+c_8\Big(\frac{2\sqrt h}{\sqrt{m_1(\varepsilon,\lambda,\delta)}}\,2^{-k}+\frac{4h}{m_1(\varepsilon,\lambda,\delta)}\Big).
\end{tcmultline*}
By~\eqref{eq:hm1}, $2\sqrt h/\sqrt{m_1(\varepsilon,\lambda,\delta)}\le\varepsilon'\lambda/\sqrt{c_9}\le\varepsilon'\lambda/c_9$ and $4h/m_1(\varepsilon,\lambda,\delta)\le(\varepsilon'\lambda)^2/c_9$, since $c_9<1$. The right side is therefore at most $(c_7+c_8/c_9)\big((\varepsilon'\lambda)^2+\varepsilon'\lambda\cdot2^{-k}\big)=c_{10}\big((\varepsilon'\lambda)^2+\varepsilon'\lambda\cdot2^{-k}\big)$, which is~\eqref{eq:requirement}.
\end{proof}
\section{Proof of Lemma~\ref{lem:gamma2contract}}
\label{app:gamma2}

This appendix proves Lemma~\ref{lem:gamma2contract}. We use the distance from a point to a nonempty finite set defined in~\eqref{eq:setdist}, and define $\mathsf d'(\cdot,\cdot)$ in the same way. We may assume $\gamma_2(\mathcal T,\mathsf d)<\infty$; otherwise the bound is trivial. For an admissible sequence $\{\mathcal T_r\}_{r\ge0}$ for $(\mathcal T,\mathsf d)$, write
\[
\Phi(\{\mathcal T_r\}):=\sup_{\mathbf x\in\mathcal T}\sum_{r\ge0}2^{r/2}\mathsf d(\mathbf x,\mathcal T_r),
\]
so that $\gamma_2(\mathcal T,\mathsf d)$ is the infimum of $\Phi$ over admissible sequences. Define $\Phi'$ in the same way on $(\varphi(\mathcal T),\mathsf d')$. The proof maps an admissible sequence for $(\mathcal T,\mathsf d)$ to its image under $\varphi$ and compares the two chaining sums term by term. Fix an admissible sequence $\{\mathcal T_r\}$ for $(\mathcal T,\mathsf d)$, and set $\mathcal T'_r:=\varphi(\mathcal T_r)$. The proof has four steps.

\emph{Step 1.} Section~\ref{sec:tools} requires $\mathcal S_r\subset\varphi(\mathcal T)$, $|\mathcal S_0|=1$ and $|\mathcal S_r|\le2^{2^r}$. We verify these conditions for $\mathcal S_r=\mathcal T'_r$. First, $\mathcal T_r\subset\mathcal T$ gives $\mathcal T'_r\subset\varphi(\mathcal T)$. Second, the image of a finite set under a map has at most as many elements as the set. Hence $|\mathcal T'_r|\le|\mathcal T_r|\le2^{2^r}$, and $|\mathcal T'_0|=1$ because $\mathcal T_0$ is a single point.

\emph{Step 2.} Fix $\mathbf x\in\mathcal T$ and $r\ge0$. For every $\mathbf t\in\mathcal T_r$,
\[
\mathsf d'\big(\varphi(\mathbf x),\mathcal T'_r\big)\lestep{a}\mathsf d'\big(\varphi(\mathbf x),\varphi(\mathbf t)\big)\lestep{b}\omega\,\mathsf d(\mathbf x,\mathbf t).
\]
Step~(a) is due to \eqref{eq:setdist}, since $\varphi(\mathbf t)\in\mathcal T'_r$. Step~(b) holds by the Lipschitz condition on $\varphi$. The left side does not depend on $\mathbf t$. Minimizing the right side over $\mathbf t\in\mathcal T_r$, and using $\omega\ge0$ and \eqref{eq:setdist}, gives
\begin{equation}
\mathsf d'\big(\varphi(\mathbf x),\mathcal T'_r\big)\le\omega\,\mathsf d(\mathbf x,\mathcal T_r)\qquad\text{for all }\mathbf x\in\mathcal T,\ r\ge0. \label{eq:app-contract}
\end{equation}

\emph{Step 3.} Fix $\mathbf z\in\varphi(\mathcal T)$, and choose $\mathbf x\in\mathcal T$ with $\varphi(\mathbf x)=\mathbf z$. Then
\begin{tcmultline*}
\sum_{r\ge0}2^{r/2}\mathsf d'(\mathbf z,\mathcal T'_r)\eqstep{a}\sum_{r\ge0}2^{r/2}\mathsf d'\big(\varphi(\mathbf x),\mathcal T'_r\big) \tcbr
\lestep{b}\omega\sum_{r\ge0}2^{r/2}\mathsf d(\mathbf x,\mathcal T_r)\lestep{c}\omega\,\Phi(\{\mathcal T_r\}).
\end{tcmultline*}
Step~(a) substitutes $\mathbf z=\varphi(\mathbf x)$. Step~(b) multiplies \eqref{eq:app-contract} by $2^{r/2}>0$ and sums over $r\ge0$; all terms are nonnegative. Step~(c) holds because $\Phi(\{\mathcal T_r\})$ is the supremum of the middle sum over $\mathbf x\in\mathcal T$. The final bound does not depend on $\mathbf z$. Taking the supremum over $\mathbf z\in\varphi(\mathcal T)$ on the left therefore gives
\begin{equation}
\Phi'(\{\mathcal T'_r\})\le\omega\,\Phi(\{\mathcal T_r\}). \label{eq:app-phi}
\end{equation}
\emph{Step 4.} By Step~1, $\{\mathcal T'_r\}$ is an admissible sequence for $(\varphi(\mathcal T),\mathsf d')$. Hence
\[
\gamma_2\big(\varphi(\mathcal T),\mathsf d'\big)\lestep{a}\Phi'(\{\mathcal T'_r\})\lestep{b}\omega\,\Phi(\{\mathcal T_r\}).
\]
Step~(a) holds because $\gamma_2(\varphi(\mathcal T),\mathsf d')$ is the infimum of $\Phi'$ over all the admissible sequences for $(\varphi(\mathcal T),\mathsf d')$. Step~(b) is \eqref{eq:app-phi}. The left side does not depend on $\{\mathcal T_r\}$, and $\{\mathcal T_r\}$ was an arbitrary admissible sequence for $(\mathcal T,\mathsf d)$. The left side is therefore a lower bound for $\omega\,\Phi(\{\mathcal T_r\})$ over all such admissible sequences, and so it is at most their infimum, $\omega\,\gamma_2(\mathcal T,\mathsf d)$. This proves $\gamma_2(\varphi(\mathcal T),\mathsf d')\le\omega\,\gamma_2(\mathcal T,\mathsf d)$.
\bibliographystyle{IEEEtran}
\bibliography{IEEEabrv,refs}

\end{document}